\PassOptionsToPackage{table,xcdraw}{xcolor}
\documentclass[acmsmall]{acmart}

\AtBeginDocument{%
  }

\usepackage{multicol, multirow}
\usepackage{subcaption}
\usepackage{color}
\usepackage{enumitem}
\usepackage{makecell}
\usepackage{pifont} 
 
\usepackage{tikz}
\usepackage{collcell}
\usepackage{siunitx}

\usepackage{graphicx}
\usepackage{tabularx}
\usepackage{float}

\usepackage{amsthm}
\usepackage{mathtools}  

\usepackage{tcolorbox}
\usepackage{xspace}
\usepackage{fontawesome}
\usepackage{setspace}

\newcommand{\xmark}{\ding{55}}%
\newcommand{\cmark}{\ding{51}}%

\newcommand*{\DanNumber}{0.4}%
\newcommand*{\MinNumber}{0.5}%
\newcommand*{\MidNumber}{0.75}%
\newcommand*{\MaxNumber}{1.0}%

\definecolor{myorange}{RGB}{255, 127, 15}
\definecolor{myblue}{RGB}{31, 119, 180}

\def\thickhline{\noalign{\hrule height 1pt}}

\newcommand{\ApplyGradient}[1]{%
        \IfDecimal{#1}{
            \ifdim #1 pt > \MidNumber pt
                \pgfmathparse{(#1-\MidNumber)/(\MaxNumber-\MidNumber+0.001)*100}%
                \global\let\colorratio\pgfmathresult
                \cellcolor{blue!\colorratio}%
                \ifdim #1 pt > 0.85 pt
                    \textcolor{white}{#1}
                \else
                    \textcolor{black}{#1}
                \fi
            \else
                \ifdim #1 pt > \MinNumber pt
                    \pgfmathparse{(#1-\MinNumber)/(\MaxNumber*2)*100}%
                    \global\let\colorratio\pgfmathresult
                    \cellcolor{blue!\colorratio}%
                    \textcolor{black}{#1}
                \else
                    \ifdim #1 pt < \DanNumber pt
                        \pgfmathparse{(\DanNumber-#1)/(\DanNumber)*100}%
                        \global\let\colorratio\pgfmathresult
                        \cellcolor{red!\colorratio}%
                        \textcolor{black}{#1}
                    \else
                        \cellcolor{white}%
                        \textcolor{black}{#1}
                    \fi
                \fi
            \fi
        }{
            \cellcolor{white}%
            \textcolor{black}{#1}
        }
}

\newcolumntype{C}{>{\collectcell\ApplyGradient}c<{\endcollectcell}}

\newtheorem{dfn}{Definition}

\newcommand{\revised}[1]{#1}
\newcommand{\problemtitle}[1]{\gdef\@problemtitle{#1}}
\newcommand{\probleminput}[1]{\gdef\@probleminput{#1}}
\newcommand{\problemquestion}[1]{\gdef\@problemquestion{#1}}
\NewEnviron{problem}{
  \problemtitle{}\probleminput{}\problemquestion{}
  \BODY
  \par\addvspace{.5\baselineskip}
  \noindent
  \begin{tabularx}{\linewidth}{@{\hspace{\parindent}} l X c}
    \multicolumn{2}{@{\hspace{\parindent}}l}{\@problemtitle} \\
    \textbf{Input:} & \@probleminput \\
    \textbf{Question:} & \@problemquestion
  \end{tabularx}
  \par\addvspace{.5\baselineskip}
}

\newcounter{probcount}

\NewTColorBox{custombox}{o}{
    colback=gray!10,
	colframe=gray!10,
	left=1.0mm,
	right=1.0mm,
	top=0.5mm,
	bottom=0.5mm,
	fonttitle=\bfseries,
	arc=0mm,
	leftrule=0mm,
	rightrule=0mm,
	toprule=0mm,
	bottomrule=0mm,
	notitle,
	before=\par\medskip\noindent,
    IfValueTF={#1}{before upper={\textbf{#1: } }}{},
    parbox=false,
}

\setcopyright{cc}
\setcctype{by-sa}
\acmDOI{10.1145/3729376}
\acmYear{2025}
\acmJournal{PACMSE}
\acmVolume{2}
\acmNumber{FSE}
\acmArticle{FSE106}
\acmMonth{7}
\received{2025-02-26}
\received[accepted]{2025-04-01}

\begin{document}

\title{Automated Trustworthiness Oracle Generation for Machine Learning Text Classifiers}

\author{Lam Nguyen Tung}
\orcid{0009-0000-3038-8403}
\affiliation{%
  \institution{Monash University}
  \city{Melbourne}
  \country{Australia}
}
\email{lam.nguyentung@monash.edu}

\author{Steven Cho}
\orcid{0009-0001-2548-4406}
\affiliation{%
  \institution{University of Auckland}
  \city{Auckland}
  \country{New Zealand}
}
\email{scho518@aucklanduni.ac.nz}

\author{Xiaoning Du}
\orcid{0000-0003-3728-9541}
\affiliation{%
  \institution{Monash University}
  \city{Melbourne}
  \country{Australia}
}
\email{xiaoning.du@monash.edu}

\author{Neelofar Neelofar}
\orcid{0000-0003-2572-0250}
\affiliation{%
  \institution{Royal Melbourne Institute of Technology}
  \city{Melbourne}
  \country{Australia}
}
\email{neelofar.neelofar@rmit.edu.au}

\author{Valerio Terragni}
\orcid{0000-0001-5885-9297}
\affiliation{%
  \institution{University of Auckland}
  \city{Auckland}
  \country{New Zealand}
}
\email{v.terragni@auckland.ac.nz}

\author{Stefano Ruberto}
\orcid{0000-0001-8666-2782}
\affiliation{%
  \institution{Joint Research Centre at the European Commission}
  \city{Ispra}
  \country{Italy}
}
\email{Stefano.RUBERTO@ec.europa.eu}

\author{Aldeida Aleti}
\orcid{0000-0002-1716-690X}
\affiliation{%
  \institution{Monash University}
  \city{Melbourne}
  \country{Australia}
}
\email{aldeida.aleti@monash.edu}

\renewcommand{\shortauthors}{Lam et al.}

\begin{abstract}
Machine learning (ML) for text classification has been widely used in various domains, such as toxicity detection, chatbot consulting, and review analysis.
These applications can significantly impact ethics, economics, and human behavior, raising serious concerns about trusting ML decisions.
Several studies indicate that traditional uncertainty metrics, such as model confidence, and performance metrics, like accuracy, are insufficient to build human trust in ML models.
These models often learn spurious correlations during training and predict based on them during inference.
When deployed in the real world, where such correlations are absent, their performance can deteriorate significantly.
To avoid this, a common practice is to test whether predictions are made reasonably based on valid patterns in the data. 
Along with this, a challenge known as the \textit{trustworthiness oracle problem} has been introduced.
So far, due to the lack of automated trustworthiness oracles, the assessment requires manual validation, based on the decision process disclosed by explanation methods. 
However, this approach is time-consuming, error-prone, and not scalable.

To address this problem, we propose TOKI, the first automated trustworthiness oracle generation method for text classifiers.
TOKI automatically checks whether the words contributing the most to a prediction are semantically related to the predicted class.
Specifically, we leverage ML explanation methods to extract the decision-contributing words and measure their semantic relatedness with the class based on word embeddings.
As a demonstration of its practical usefulness, we also introduce a novel adversarial attack method that targets trustworthiness vulnerabilities identified by TOKI.
We compare TOKI with a naive baseline based solely on model confidence. 
To evaluate their alignment with human judgement, experiments are conducted on human-created ground truths of approximately 8,000 predictions.
Additionally, we compare the effectiveness of TOKI-guided adversarial attack method with A2T, a state-of-the-art adversarial attack method for text classification. 
Results show that 
(1) relying on prediction uncertainty metrics, such as model confidence, cannot effectively distinguish between trustworthy and untrustworthy predictions,
(2) TOKI achieves 142\% higher accuracy than the naive baseline, and
(3) TOKI-guided adversarial attack method is more effective with fewer perturbations than A2T.
\end{abstract}

\ccsdesc[500]{Software and its engineering~Software testing and debugging}
\ccsdesc[500]{Human-centered computing}
\ccsdesc[300]{Computing methodologies~Natural language processing}
\ccsdesc[100]{Computing methodologies~Learning paradigms}

\keywords{SE4AI, Oracle Problem, Trustworthy Text Classifier, Explainability
}


\maketitle

\textcolor{red}{\small\textit{Warning: This paper contains examples of language that some people may find offensive or upsetting.
}}

\section{Introduction}
\label{sec:introduction}
Machine learning (ML) 
holds significant importance
in contemporary advanced systems, such as spam detection, clinical text analysis, and vulnerability detection, with text classification being a primary application. Despite their superior performance during development, ML models can still fail in real-life scenarios~\cite{Caruana:2015:IntelligibleModelsforHealthCarePredictingPneumoniaRiskandHospital30dayReadmission, Lapuschkin:2019:UnmaskingCleverHansPredictorsAndAssessingWhatMachinesReallyLearn}, 
raising concerns about trusting their decisions. 
When assessing a model, it is important not only to evaluate its general task-solving ability using metrics, such as prediction uncertainty or classification accuracy, but also to understand its decision-making process~\cite{Liao:2023:AITransparencyintheAgeofLLMsAHumanCenteredResearchRoadmap}. 
Indeed, several studies show that these metrics alone are insufficient indicators of model reliability~\cite{Anh:2015:DeepNeuralNetworksAreEasilyFooledHighConfidencePredictionsForUnrecognizableImages, Canbek:2022:PToPIA:ComprehensiveReviewAnalysisandKnowledgeRepresentationofBinaryClassificationPerformanceMeasuresMetrics},
\begin{figure}[t]
    \centering
    \begin{subfigure}[t]{0.495\linewidth}
        \centering
        \includegraphics[width=1.\linewidth, trim={0 25.4cm 0 3mm}, clip]{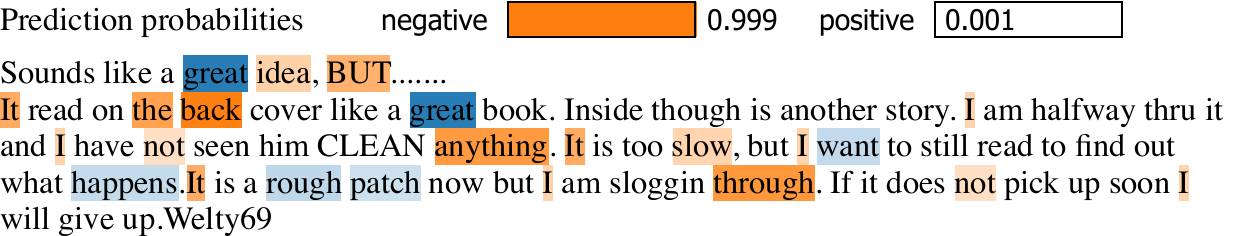}
        \vspace{-5mm}
        \caption{Original text}
        \label{fig:amazon_example}
    \end{subfigure}%
    \hfill
    \begin{subfigure}[t]{0.495\linewidth}
        \centering
        \includegraphics[width=1.\linewidth, trim={0 25.4cm 0 3mm}, clip]{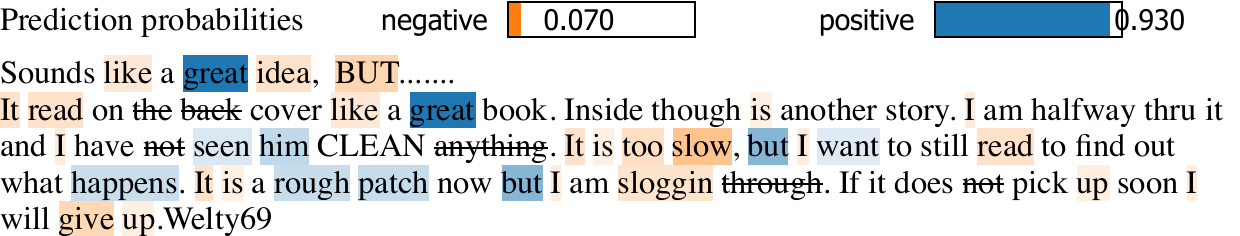}
        \vspace{-5mm}
        \caption{Some irrelevant words (strikethrough) removed}
        \label{fig:amazon_remove_words}
    \end{subfigure}%
    \hfill
    \begin{subfigure}[t]{0.495\linewidth}
        \centering
        \includegraphics[width=1.\linewidth, trim={0 24.8cm 0 -5mm}, clip]{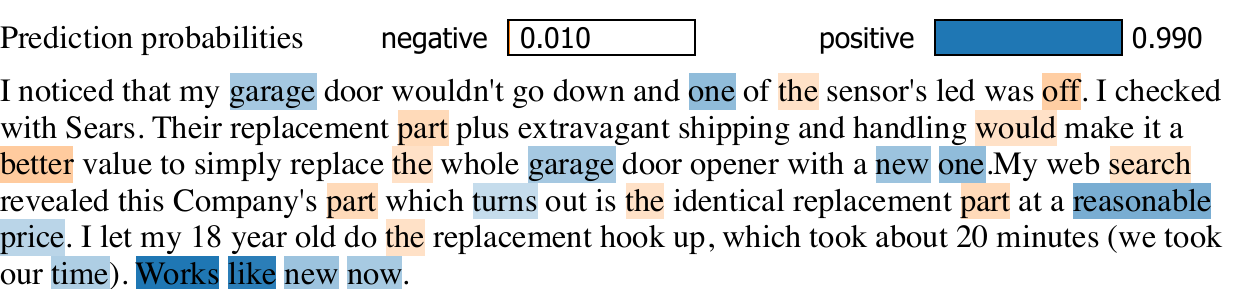}
        \vspace{-5mm}
        \caption{A trustworthy prediction of a positive review}
        \label{fig:original_trust_prediction}
    \end{subfigure}%
    \hfill
    \begin{subfigure}[t]{0.495\linewidth}
        \centering
        \includegraphics[width=1.\linewidth, trim={0 24.8cm 0 -5mm}, clip]{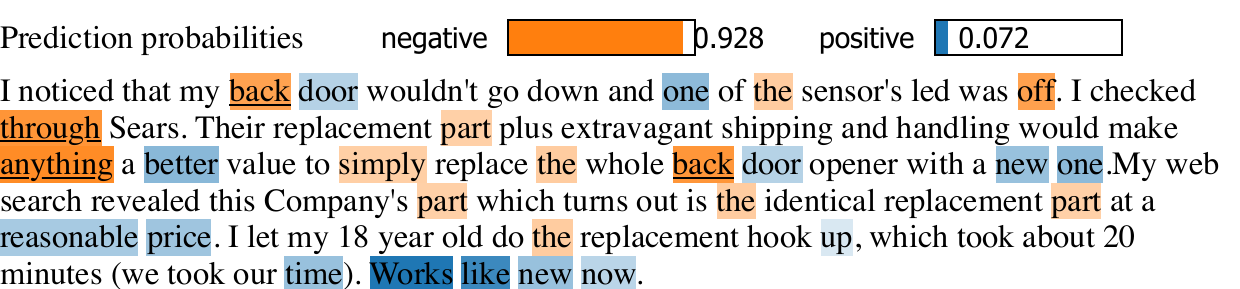}
        \vspace{-5mm}
        \caption{Irrelevant words (underlined) injected}
        \label{fig:falied_synthesised_data}
    \end{subfigure}%
    \vspace{-2mm}
    \caption{LIME explanations of predictions analysing amazon reviews.
    Words highlighted in \textcolor{myorange}{orange} and \textcolor{myblue}{blue} indicate which class, \textcolor{myorange}{\textbf{negative}} or \textcolor{myblue}{\textbf{positive}}, they contribute to, with the shade representing their importance.
    }
    \Description[Examples for differentiating trustworthiness and robustness]{Fully described in the text.}
    \label{exp:intro}
\end{figure}
as the model might learn spurious patterns~\cite{Ye:2024:SpuriousCorrelationsinMachineLearningASurvey}, leading to overconfident decisions based on irrelevant features~\cite{Geirhos:2020:ShortcutLearninginDeepNeuralNetworks}.

An ML model relying on irrelevant features, when faced with unseen data, can misclassify in the absence of these features or be fooled by them, leading to problematic predictions.
We illustrate this using a Bert-based~\cite{Devlin::2019:BERTPretrainingofDeepBidirectionalTransformersforLanguageUnderstanding} binary sentiment classifier applied to amazon reviews.
Figure~\ref{exp:intro} shows four review examples, the model's predictions, and the probabilities of each class.
We also highlight the important words steering these predictions, which are identified by LIME~\cite{Ribeiro:2016:WhyShouldITrustYouExplainingthePredictionsofAnyClassifier}, an explanation method. 
Orange and blue colors highlight words contributing to the negative and positive classes, respectively, with darker shades indicating higher importance.
The review in Figure~\ref{fig:amazon_example} is correctly classified as negative with high confidence.
However, a closer look at the most important words the classifier relies on reveals that 
these words,
such as ``back'', ``the'', ``anything'', and ``through'', are unrelated to either class.
When removing some of them, as shown in Figure~\ref{fig:amazon_remove_words},
the review is misclassified as positive by the same model, despite negative sentiment in phrases 
``too slow'', ``rough patch'', ``sloggin'', and ``give up''.
To further explore the potential harm of this phenomenon, we synthesise a new review using a positive review shown in Figure~\ref{fig:original_trust_prediction}.
This new review, as illustrated in Figure~\ref{fig:falied_synthesised_data}, is injected with the words ``back'', ``through'', and ``anything''.
Although, these injections do not change the original meaning, the confident correct prediction is flipped to its opposite.
These shifts underscore the importance of assessing whether the learned patterns are genuinely valid and generalisable or if they are merely based on spurious correlations within the training data, typically referred to as \textit{shortcut learning}~\cite{
Du:2023:ShortcutLearningofLargeLanguageModelsinNaturalLanguageUnderstanding}.
A shortcut learning model is unlikely to provide correct classifications for the right reasons.
Therefore, it becomes ineffective once deployed in the real world, where spurious correlations are absent. 

To ensure the quality of ML systems in a reliable and cost-effective way, considerable effort has been focused on automating various aspects of the testing process, particularly through automated test oracles~\cite{Barr:2015:TheOracleProbleminSoftwareTestingASurvey}.
Traditional software testing has inspired automated test oracles for ML to test various properties, such as correctness, fairness, and robustness using techniques like metamorphic testing~\cite{Pei:2017:DeepXplore:AutomatedWhiteboxTestingofDeepLearningSystems, Guo:2018:DLFuzz:DifferentialFuzzingTestingofDeepLearningSystems}.
However, trustworthiness testing has not received as much attention and consequently remains less resolved~\cite{Cho:2024:TOWER}.
Trustworthiness refers to the ability of a model to make reasonable predictions based on relevant features, such as semantically related words in text classification.
Developing automated test oracles for trustworthiness testing is a long-lasting challenge hindering the broader adoption of ML models in the real world.
Due to the lack of automated trustworthiness oracles, most studies~\mbox{\cite{Ribeiro:2016:WhyShouldITrustYouExplainingthePredictionsofAnyClassifier, Lapuschkin:2019:UnmaskingCleverHansPredictorsAndAssessingWhatMachinesReallyLearn, Du::2021:TowardsInterpretingandMitigatingShortcutLearningBehaviorofNLUmodels}} rely on human-based evaluation as the oracle to assess the ML prediction reasoning uncovered by explanation methods~{\cite{Zhao:2024:ExplainabilityforLargeLanguageModelsASurvey}}.
However, this is time-consuming, error-prone, and not scalable~{\cite{Ye:2024:SpuriousCorrelationsinMachineLearningASurvey}}.

In this paper, we focus on the oracle problem of assessing the trustworthiness of predictions made by text classifiers.
Intuitively, trustworthiness assessment can be done by measuring the semantic relatedness between each decision-contributing word and the class name.
However, measuring the semantic relatedness between two words is non-trivial, as discussed in Section~\ref{sec:define_trustworthiness}, 
Given a word, there exists a group of words naturally related to it with very similar semantics like its synonyms.
Their semantic similarity can be easily measured with existing metrics, such as cosine similarity.
Some other words can also be related to the target word, but require a semantic hop, which we call indirectly related words.
For example, the words ``\textit{computer}'' and ``\textit{file}'' are naturally related.
Although the word ``\textit{extension}'' may not seem naturally related to ``\textit{computer}'', it is strongly semantically connected to ``\textit{file}'' and thus indirectly related to ``\textit{computer}''.
We argue that examining the distribution of words helps better recognise semantically related words.
Based on this, we propose an automated trustworthiness oracle generation method that leverages explanation methods to extract decision-contributing words and assesses their semantic relatedness to the class based on such distribution.
The key idea is to identify keywords, which act as anchors for assessing semantic relatedness and indicate what a text classifier should rely on for predictions.
A prediction is then deemed trustworthy if it is mainly based on these keywords.
For instance, the prediction in Figure~\ref{fig:amazon_example} is untrustworthy because the top contributing words, ``back'', ``the'', ``anything'', and ``through'', are all semantically unrelated to ``negative''.
To reveal the negative impact of vulnerabilities in trustworthiness, we also design an adversarial attack method guided by trustworthiness oracles.
Our main contributions are summarised as follows.
\begin{itemize}
    \item TOKI, the first 
    approach for generating automated trustworthiness oracles. 
    \item A novel attack method that targets trustworthiness vulnerabilities identified by TOKI.
    
    \item A benchmark for trustworthiness assessment for text classification, which contains approximately 8,000 predictions in various domains, such as topic classification, sentiment analysis, clinical mental text classification, hate speech detection, and software issue management.

    \item An investigation on the relation between the uncertainty and trustworthiness of predictions,
    revealing that relying on prediction uncertainty metrics, such as model confidence, overlooks untrustworthy high-confidence predictions and trustworthy low-confidence predictions. 
    
    
    \item Ablation studies and comparative evaluations of TOKI.
    For trustworthiness assessment, we compare TOKI with a naive baseline solely based on model confidence.
    For adversarial attacks, we compare our method with A2T~\cite{Yoo:2021:TowardsImprovingAdversarialTrainingofNLPModels}, a state-of-the-art (SOTA) adversarial attack method.
    The results show TOKI's superior effectiveness and efficiency.


\end{itemize}


\section{Problem Definition, Preliminary, and Motivation}
\label{sec:problem_definition}
This section clarifies the definition of trustworthiness used in the paper, the trustworthiness oracle problem, and our intuition to address it.


\subsection{Definition of Trustworthiness} \label{sec:define_trustworthiness}


Trustworthiness is a complex concept that has raised numerous scholarly debates among researchers.
This paper focuses on trustworthiness, in particular, whether a model makes predictions based on valid and reasonable patterns rather than spurious ones.
Hence, we adopt the definition of trustworthiness proposed by {\citeN{Kästner:2021:OntheRelationofTrustandExplainabilityWhytoEngineerforTrustworthiness}} as shown in Definition~{\ref{def:trustworthiness}}.  
\begin{dfn}
\label{def:trustworthiness}
An ML model is trustworthy to a stakeholder in a given context if and only if it works properly in the context and the stakeholder has justified belief in it.
\end{dfn}
\noindent 
They also distinguish trustworthiness from trust: trustworthiness is a system's property, as recognised by prior studies as a critical non-functional requirement~\cite{Riccio:2020:TestingMachineLearningBasedSystems:ASystematicMapping}, whereas trust is the perception a person has towards it.
Hence, people can still trust an untrustworthy system.

According to Definition~{\ref{def:trustworthiness}}, understanding a model thoroughly helps justify our beliefs about how well it works.
Model explanations can serve as a means to gain this understanding~{\cite{Wiegreffe:2019:AttentionIsNotNotExplanation}}, thereby promoting trustworthiness~{\cite{Bussone:2015:TheRoleofExplanationsonTrustandRelianceinClinicalDecisionSupportSystems, Kästner:2021:OntheRelationofTrustandExplainabilityWhytoEngineerforTrustworthiness}}. 
There are two main types of model explanations:  global~\cite{Caruana:2015:IntelligibleModelsforHealthCarePredictingPneumoniaRiskandHospital30dayReadmission}, which provides insights into the entire model's inner workings, and local~\cite{Ribeiro:2016:WhyShouldITrustYouExplainingthePredictionsofAnyClassifier, Li:2017:UnderstandingNeuralNetworksthroughRepresentationErasure, Mohebbi:2021:ExploringtheRoleofBERTTokenRepresentationstoExplainSentenceProbingResults}, which focuses on individual predictions.
Local explanations, by breaking down a model into its components, allow users to grasp its functionality and decision-making process in a way that aligns with human cognitive patterns.
Hence, they are more readily applicable~\cite{Adadi:2018:PeekingInsidetheBlackBoxASurveyonExplainableArtificialIntelligenceXAI}.
\revised{In text classification, various local explanations have been developed, such as feature attribution-based~\cite{Li:2017:UnderstandingNeuralNetworksthroughRepresentationErasure, Mohebbi:2021:ExploringtheRoleofBERTTokenRepresentationstoExplainSentenceProbingResults}, attention-based~\cite{Yeh:2024:AttentionViz:AGlobalViewofTransformerAttention, Barkan:2021:GradSAM:ExplainingTransformersviaGradientSelfAttentionMaps}, and counterfactual~\cite{Treviso:2023:CREST:AJointFrameworkforRationalizationandCounterfactualTextGeneration} explanations.
While other local explanations are subject to extensive debate~\cite{Jain:2019:AttentionisnotExplanation}, feature attribution-based explanations, such as 
LIME~\cite{Ribeiro:2016:WhyShouldITrustYouExplainingthePredictionsofAnyClassifier}, have proven to be effective, faithful, and widely used~\cite{Mariotti:2024:TextFocus:AssessingtheFaithfulnessofFeatureAttributionMethodsExplanationsinNaturalLanguageProcessing}.
Hence, we leverage these explanations
to uncover the reasoning behind individual predictions by highlighting the most relevant features contributing to those predictions.
Exploring alternative explanations for assessing trustworthiness is an interesting avenue
for future work.}

A trustworthy model should make correct predictions, and the reasoning behind them, supported by explanations, should also be plausible.
We propose the definition of a trustworthy prediction. 
\begin{dfn}
\label{def:trustworthy_prediction}
A trustworthy prediction is correct and the reasoning behind it is also \textbf{plausible}.
\end{dfn}
\noindent 
In the context of text classification, a trustworthy prediction should rely on words in the input text that align with human reasoning.
To simulate human judgement, 
the reasoning behind a prediction can be considered plausible if words contributing the most to the prediction are \textbf{semantically related} to the predicted class.
We follow the definition of semantic relatedness described by \citeN{Budanitsky:2006:EvaluatingWordNetbasedMeasuresofLexicalSemanticRelatedness}. 
Words can be semantically related by lexical relationships, such as meronymy (car--wheel) and antonymy (hot--cold), or just by any kind of functional association or other ``non-classical relations'' (pencil--paper, penguin--Antarctica, and rain--flood)~\cite{Morris:2004:NonClassicalLexicalSemanticRelations}.
It is important to distinguish semantic relatedness from \textbf{semantic similarity}.
Semantic relatedness is a more general concept~\cite{Budanitsky:2006:EvaluatingWordNetbasedMeasuresofLexicalSemanticRelatedness} while semantic similarity refers to the degree of overlap or resemblance in meaning between two words~\cite{Thabet:2013:DescriptionandEvaluationofSemanticSimilarityMeasuresApproaches}.
For example, ``desk'' and ``chair'' are semantically related but not semantically similar, ``desk'' and ``table'', on the other hand, are both semantically related and semantically similar.
It is also important to note that semantically relatedness is contextual, meaning two words are semantically related in a specific context, but might not in others.
For instance, ``bank'' in ``\textit{the bank exploits small firms}'' is semantically related to economics, while ``bank'' in ``\textit{we walked along the river bank}'' is not.
Definition~{\ref{def:trustworthy_prediction}} describes a trustworthy prediction, referring to \textit{local} trustworthiness.
In contrast, \textit{global} trustworthiness is the ability to make trustworthy predictions across a broad range of inputs. 
A globally trustworthy model can perfectly interpret every input like human domain experts. 
Although achieving \textit{global} trustworthiness is the ultimate goal, it is challenging due to the more complex architectures and training procedures required, which result in higher computational costs.
Therefore, we focus on \textit{local} trustworthiness of individual predictions.
While most observed predictions by a model are trustworthy, this does not directly mean the entire model is trustworthy. 


\subsection{Trustworthiness and Robustness}

Sharing the same goal of improving the model's generalisability, robustness testing might be able to identify problems that can be spotted during trustworthiness testing.
Local robustness measures the model's ability to retain its prediction on a sample under perturbations, also known as adversarial examples~\cite{Zhang:2020:AdversarialAttacksonDeeplearningModelsinNaturalLanguageProcessingASurvey}, that do not affect human perception and decision.
For example, replacing a few words in Figure~\ref{fig:amazon_example}  with their synonyms should not alter a human’s decision to classify it as negative.
If the model fails to maintain its original prediction on this adversarial example, we can conclude that the adversarial example reveals a local robustness issue.

Before investigating the oracle problem of model trustworthiness, one important question we need to answer is that to what extent does the trustworthiness problem differ from the robustness problem?
More essentially, are there any issues that can be uncovered during trustworthiness testing but not during robustness testing?
When a model's prediction is trustworthy on a sample, the prediction is correct and based on justifiable reasons.
This prediction is not necessarily robust, meaning it might be altered under minor perturbations, such as replacing the decision-essential words with their semantic equivalents, or removing/modifying the decision-unessential ones.
On the other hand, when a prediction by the model is robust, it is still possible that the model relies on some justifiable shortcut reasons, resulting in problematic predictions in the future where the shortcuts are absent or the shortcuts appear in samples having opposite classes.
Hence, we conclude that the trustworthiness problem overlaps with the robustness problem, but they are not the same.

\begin{figure}[t]
    \centering
    \begin{subfigure}[ht]{0.495\linewidth}
        \centering
        \includegraphics[width=1.\linewidth, trim={0 25.4cm 0 3mm}, clip]{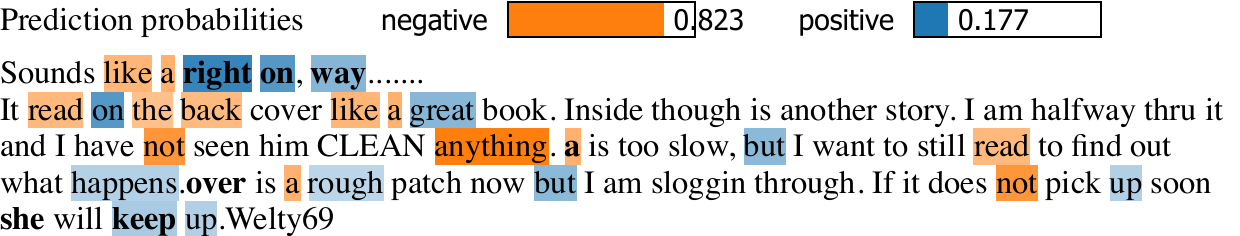}
        \vspace{-5mm}
        \caption{An adversarial example generated by TextAttack}
        \label{fig:bae_adversarial_attack}
    \end{subfigure}%
    \hfill
    \begin{subfigure}[ht]{0.49\linewidth}
        \centering
        \includegraphics[width=1.\linewidth, trim={0 25.4cm 0 3mm}, clip]{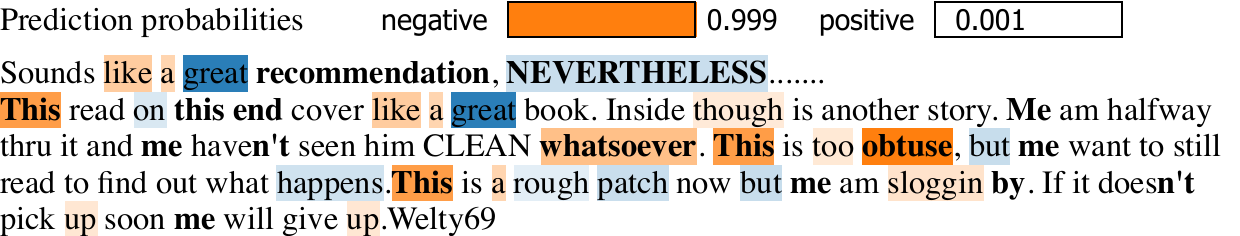}
        \vspace{-5mm}
        \caption{An adversarial example generated manually}
        \label{fig:man_adversarial_attack}
    \end{subfigure}%
    \vspace{-2mm}
    \caption{LIME explanations of predictions analysing amazon reviews.
    \textbf{Bold} words represent perturbations, while words highlighted in \textcolor{myorange}{orange} and \textcolor{myblue}{blue} indicate their contribution to the \textcolor{myorange}{\textbf{negative}} or \textcolor{myblue}{\textbf{positive}} class, with the shade reflecting their importance.
    }
    \Description[Examples for differentiating trustworthiness and robustness]{Fully described in the text.}
\end{figure}
We differentiate trustworthiness from robustness by revisiting the prediction in Figure~\ref{fig:amazon_example}, showing that robustness testing is unable to identify issues in the model's behavior, but trustworthiness testing can.
TextAttack~\cite{Morris:2020:TextAttack:AFrameworkforAdversarialAttacksDataAugmentationandAdversarialTraininginNLP},
a framework implementing several SOTA adversarial example generation methods, is applied to attack~\cite{Zhang:2020:AdversarialAttacksonDeeplearningModelsinNaturalLanguageProcessingASurvey} the model on the same text input.
An additional adversarial example is generated manually to preserve grammar and semantic by replacing orange-highlighted words in Figure~\ref{fig:amazon_example} with their synonyms from WordNet~\cite{Miller:1995:WordNetALexicalDatabaseforEnglish}.
The substitutions are \ding{172}~idea--recommendation, \ding{173}~BUT--NEVERTHELESS, \ding{174}~It--This, \ding{175}~the--this, \ding{176}~back--end, \ding{177}~anything--whatsoever, \ding{178}~slow--obtuse, \ding{179}~I--me, \ding{180}~through--by, \ding{181}~not--n't.
Figures~\ref{fig:bae_adversarial_attack} and~\ref{fig:man_adversarial_attack}, with bold words representing perturbations, show LIME explanations for two predictions to adversarial examples generated by TextAttack and manually, respectively. 
All adversarial examples 
fail to attack the model since it maintains the same prediction.
This indicates that although the model is robust against certain adversarial examples, its prediction can still be untrustworthy.

\subsection{Trustworthiness Oracle} \label{sec:trustworthiness_oracle_problem}

ML testing involves providing a model with inputs and observing its responses.
The oracle problem
in ML testing is the challenge of determining whether these responses are appropriate.
In trustworthiness testing, a response is an explainable prediction, as described in Definition~\ref{def:explainable_prediction}.
\begin{dfn}
\label{def:explainable_prediction}
For the classifier $m$, $x$ is an individual data input into $m$, and $p = \langle \hat{y}, e \rangle$ is the \textbf{explainable prediction} to $x$ of $m$, where $\hat{y}$ is the predicted class and $e$ is the explanation.
\end{dfn}
\noindent
The explainable prediction $p$ assigns the input $x$ to the predicted class $\hat{y}$, with its reasoning explained by the explanation $e$.
The explanation $e$ is a list of decision-contributing words and the corresponding \textbf{importance scores} measuring their contribution to the prediction $p$.

Three cases can occur with an explainable prediction: (1) incorrect, (2) correct due to semantically related words, and (3) correct due to semantically unrelated words.
In the first case, investigating incorrectness becomes more important than trustworthiness assessment. 
Although the explanation might hint at why the prediction is incorrect, we leave this interesting avenue for future work. 
In this way, this paper only focuses on 
differentiating 
the last two cases, 
meaning that it examines only \textit{correct} explainable predictions in the context of trustworthiness testing.
We then adopt the definition of test oracles from \citeN{Barr:2015:TheOracleProbleminSoftwareTestingASurvey}. Definition~\ref{def:trustworthiness_test_activities} describes trustworthiness test data while Definition~\ref{def:trustworthiness_oracle} defines trustworthiness oracles.
\begin{dfn}
\label{def:trustworthiness_test_activities}
For the classifier $m$, X is the dataset correctly predicted by $m$ and P is the set of correct explainable predictions to an instance of $m$. Trustworthiness test data forms the set $T = X \uplus P$.
\end{dfn}
\begin{dfn}
\label{def:trustworthiness_oracle}
A trustworthiness oracle $D: T \mapsto \mathbb{B}$ is a 
function from an instance of trustworthiness test data $t$ to true or false, indicating whether the prediction p in t is trustworthy according to Definition~\ref{def:trustworthy_prediction}.
\end{dfn}
\noindent
A trustworthiness oracle is a predicate that determines whether the prediction in an individual trustworthiness test data is trustworthy according to Definition~\ref{def:trustworthy_prediction}. 
The trustworthiness oracle formulated in 
Definition~\ref{def:trustworthiness_oracle} can be applied to any text classifier, as long as there is an explanation method that uncovers the reasoning behind its predictions in the form of word attributions.

\subsection{Applications of Trustworthiness Oracles}

The trustworthiness problem exists in various ML applications. 
For example, a clinical mental text classifier from social media posts shows good held-out performance but might rely on irrelevant clues
frequently present in the training data, such as post tags, job positions, and special occasions~\cite{Harrigian:2020:DoModelsofMentalHealthBasedonSocialMediaDataGeneralize}. When deployed in the real world, such classifier is unlikely to perform well on unseen data.
Hence, it is crucial to assess the reasoning of ML models to detect any behavioral issues before deploying them in real-world scenarios, rather than only evaluating their classification performance using metrics such as accuracy, recall, and F1-score.

\begin{figure}[t]
    \centering
    \label{fig:ML_dev_process}
    \begin{subfigure}[ht]{0.45\linewidth}
        \centering
        \includegraphics[scale=0.50, trim={0 2mm 0 0mm}, clip]{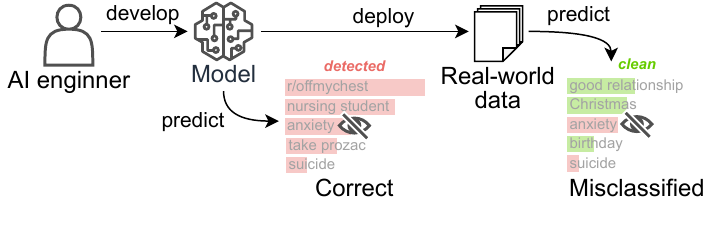}
        \vspace{-2mm}
        \caption{Traditional development process}
        \label{fig:tra_ML_dev_process}
    \end{subfigure}%
    \hfill
    \begin{subfigure}[ht]{0.55\linewidth}
        \centering
        \includegraphics[scale=0.50, trim={0 4mm 0 0mm}, clip]{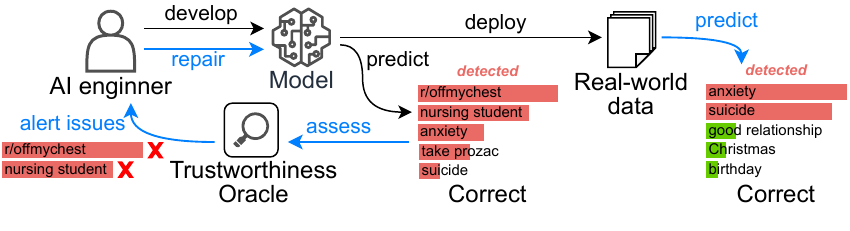}
        \vspace{-2mm}
        \caption{Integration with trustworthiness testing}
        \label{fig:new_ML_dev_process}
    \end{subfigure}%
    \vspace{-2mm}
    \caption{Comparison between two ML system development processes: without and with trustworthiness testing. \textcolor{blue}{Blue} lines indicate the differences between these processes.}
    \Description{Comparison between the development process with and without trustworthiness testing.}
\end{figure}

Figure~\ref{fig:tra_ML_dev_process} shows a traditional development process of ML systems. 
AI engineers collect all available data, preprocess and split it into training and test sets. 
The training data is used to train the model, while the test data evaluates its performance. 
If the model performs poorly, the engineers improve the learning algorithm to enhance performance. 
Once the model achieves good held-out performance, it can be deployed in the real world.
However, its performance often deteriorates due to the model's reliance on spurious correlations learned during training, making it overconfident in held-out evaluations. 
In the real world, where these correlations are absent, the model tends to misclassify unseen data.
A common solution is to augment the dataset with new manually annotated data and retrain the model, which is costly and inefficient.

Figure~\ref{fig:new_ML_dev_process} shows the process integrated with trustworthiness testing.
In addition to evaluating traditional performance metrics, this process also assesses the model's behavior behind correct predictions. 
In this process, trustworthiness oracles determine whether a prediction is trustworthy or based on spurious correlations.
Identifying such issues allows the engineers to mitigate the impact of spurious correlations, ensure the model is correct for the right reasons, and improve its generalizability.
If undetected, these issues can lead to problematic predictions in the future.
Human annotations often serve as trustworthiness oracles, but this approach is not scalable~\cite{Ye:2024:SpuriousCorrelationsinMachineLearningASurvey}. 
\revised{
Integrating automated trustworthiness oracles into the software engineering (SE) lifecycle for ML systems can greatly advance their development and applications.
Automated trustworthiness oracles enable the automation of trustworthiness testing, which is vital and closely intertwined with other SE 
activities~\cite{
Riccio:2020:TestingMachineLearningBasedSystems:ASystematicMapping}. 
This is especially valuable in the iterative development of ML systems where performance and trustworthiness must be evaluated and refined~\cite{Silverio:2022:SoftwareEngineeringforAIBasedSystemsASurvey}. 
Specifically, automated trustworthiness oracles support continuous integration and delivery pipelines by automating trustworthiness testing for ML systems.
They also serve as a
tool for monitoring ML systems in real-world environments, particularly in online testing or DevOps workflows, to verify whether predictions are trustworthy in real time, and test with real-world and corner-cased inputs.
Moreover, the outputs of trustworthiness oracles provide actionable insights for feedback-driven repairing and improving processes, 
reducing the need for human intervention.
}

\revised{
In addition, trustworthiness oracles help assess the trustworthiness of systems in SE that employ ML for text classification.
ML has been widely used in SE for text classification, automating numerous tasks to enhance planning, design, and maintenance.
For example, sentiment analysis has been applied to various SE artifacts, including git commit comments~\cite{Sinha:2016:AnalyzingDeveloperSentimentInCommitLogs}, JIRA issues~\cite{Ortu:2015:AreBulliesMoreProductiveEmpiricalStudyofAffectivenessvsIssueFixingTime}, and apps’ reviews~\cite{Panichella:2015:HowCanIImproveMyAppClassifyingUserReviewsForSoftwareMaintenanceAndEvolution}.
It also helps assess developers' psychological states~\cite{Guzman2013:TowardsEmotionalAwarenessInSoftwareDevelopmentTeams}, and analyse sentiment on Q\&A sites like StackOverflow to recommend improvements for source code~\cite{Rahman:2015:RecommendingInsightfulCommentsForSourceCodeUsingCrowdsourcedKnowledge} or to identify problematic API design features~\cite{Zhang:2013:ExtractingProblematicApiFeaturesFromForumDiscussions}.
In addition to sentiment analysis, other tasks, such as software requirement classification~\cite{Pérez:2020:ASystematicLiteratureReviewonMachineLearningforAutomatedRequirementsClassification} and project issue categorisation~\cite{Schulte:2024:StudyingTheExplanationsForTheAutomatedPredictionOfBugAndNonbugIssuesUsingLimeAndShap}, have also been integrated with ML to reduce human effort and improve efficiency.
Hence, automated trustworthiness oracles offer opportunities to foster greater trust in these systems, both within SE domains and more broadly.}

\section{TOKI: Trustworthiness Oracle through Keyword Identification}
\label{sec:approaches}

\begin{figure*}[t]
    \centering
    \includegraphics[width=\linewidth, clip, trim=2mm 0 3mm 0]{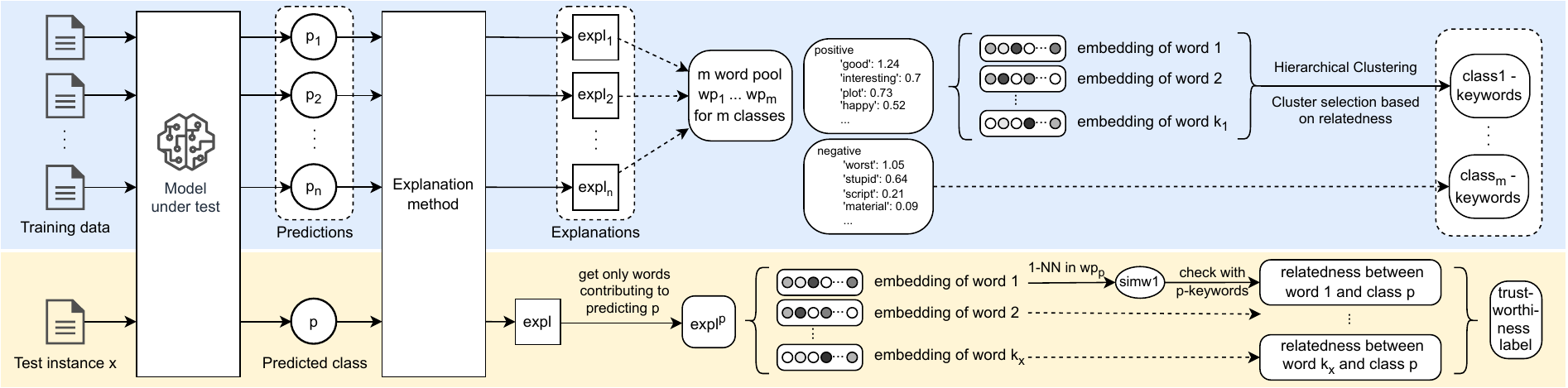}
    \caption{The process of TOKI, with blue and yellow background colors indicating the first and second pipelines.}
    \label{fig:new_approach}
    \Description[The process of TOKI, including two pipelines]{Fully described in the text.}
\end{figure*}

As outlined in Section~\ref{sec:problem_definition}, a prediction is considered trustworthy if its reasoning is plausible.
In text classification, a trustworthy prediction should rely on words semantically related to the predicted class.
We argue that examining the distribution of words helps better recognise these semantically related words.
Specifically, directly related words can act as anchors, and words indirectly related to the class are likely to form clusters around them. 
Following this intuition, we present \textit{\textbf{T}rustworthiness \textbf{O}racle through \textbf{K}eyword \textbf{I}dentification} (TOKI).
The key idea of TOKI is to use a list of \textbf{\textit{keywords}} for each class to indicate what a text classifier should rely on for predictions.
To identify these keywords, TOKI selects clusters containing directly related words and their surrounding indirectly related words.
A prediction is then deemed trustworthy if it mainly relies on the keywords of the predicted class.
While the ultimate goal is to identify a complete list of keywords, this is impractical due to resource and computational constraints.
Hence, we identify only a partial list of keywords by extracting decision-contributing words from the classifier's responses on the training data and applying clustering analysis to them.
Since not all decision-contributing words are genuinely related to the class, clustering analysis also helps separate related words from unrelated ones.
Figure~\ref{fig:new_approach} describes TOKI with two pipelines: keyword identification and trustworthiness label computation.

\label{sec:toki}

\subsection{Keyword Identification}
ML models learn correlations in training data, including both valid and spurious ones~\cite{Geirhos:2020:ShortcutLearninginDeepNeuralNetworks}.
Therefore, \textit{not all predictions are solely based on the spurious correlations}.
Predictions reflecting the valid correlations rely on \textit{words semantically related to the predicted class, regardless of their importance scores}.
While the spurious correlations can make predictions untrustworthy, TOKI identifies keywords for each class through the valid correlations, following four steps. 


\textit{\textbf{Step 1: Explaining}.}
To extract keywords, we focus on analysing the reasoning behind correct predictions, as incorrect ones might provide meaningless reasoning.
Therefore, TOKI identifies the decision-contributing words from instances that have been predicted correctly in the training set. 
Let $X$, $f(\cdot)$, and $c$ represent the training set, the text classifier, and the true class, respectively.
The explaining step is formalised in Equation~\ref{eqn:explaining}, which returns a list of explanations $E_c$ for each class $c$. 
\begin{equation}\label{eqn:explaining}
    E_c = \bigl\{ e(f, x) \mid x \in X, f(x) = c \bigr\}
\end{equation}
TOKI leverages the explanation method $e(\cdot, \cdot)$ to measure the contribution $s$ of each word $w$ to a prediction.
Several methods, such as LIME~\cite{Ribeiro:2016:WhyShouldITrustYouExplainingthePredictionsofAnyClassifier}, achieve this by locally approximating the model as an interpretable surrogate model.
Other methods~\cite{Li:2017:UnderstandingNeuralNetworksthroughRepresentationErasure} perturb input and evaluate model output changes.
Another common way~\cite{Mohebbi:2021:ExploringtheRoleofBERTTokenRepresentationstoExplainSentenceProbingResults} is to compute the gradient of the output with respect to the input.
These methods explain the prediction $f(x)$ in the form of $e(f, x) = \{\langle w, s \rangle\}$, a list of top decision-contributing words with their importance scores. 

\textit{\textbf{Step 2: Word pool construction}.}
A word pool of decision-contributing words and their averaged importance scores across explanations is created for each class, as formalised in Equation~\ref{eqn:wordpool}.
\begin{equation}\label{eqn:wordpool}
    W_c = \bigl\{ \langle w, \overline{s_w} \rangle \mid w \in E_c \bigr\},\;\text{where}\;\;
    \overline{s_w} = 
    \frac{\sum_{e \in E_c, \langle w, s_i \rangle \in e}s_i}{\sum_{e \in E_c, \langle w, s_i \rangle \in e}1} 
\end{equation}
Words in the explanations $E_c$ are categorised based on the predicted class $c$, with their importance scores averaged. 
This results in a word pool $W_c = \{\langle w, \overline{s_w} \rangle\}$ for the class $c$, containing words and their averaged importance scores.
The averaged importance score $\overline{s_w}$ also indicates the \textbf{\textit{correlation}} between the word $w$ and the class $c$.

\textit{\textbf{Step 3: Word clustering}.} 
The word pool $W_c$ of the class $c$ contains both keywords and unrelated words.
To distinguish them and collect both directly related and indirectly related words, TOKI clusters the word pool $W_c$, as formalised in Equation~\ref{eqn:clustering}. 
\begin{equation}\label{eqn:clustering}
    C_c = \text{\textit{hierarchical\_cluster}}\bigl(W_c, \theta\textsubscript{\textit{dist}} \bigr)
\end{equation}
Words in $W_c$ are transformed into embeddings by embedding methods.
As the number of clusters is unknown,
hierarchical clustering~\cite{Nielsen:2016:HierarchicalClustering} is applied to group these word embeddings based on their cosine similarities.
Clusters $C_c$ of the class $c$ are obtained by cutting the dendrogram, a hierarchical tree of relationships between the word embeddings, at the threshold distance $\theta$\textsubscript{\textit{dist}}. 

\textit{\textbf{Step 4: Keyword selection}.} 
The list of keywords is identified by selecting the clusters of words semantically related to the class name, as described in Equation~\ref{eqn:selection}.
\begin{equation}\label{eqn:selection}
\begin{aligned}
    K_c = \bigcup_{C_{i} \in C_c} \Bigl\{ C_{i} \mid sim\bigl(\overline{C_{i}^w}, c\bigr) \ge \theta\textsubscript{\textit{relate}} \Bigr\}
    ,\;\text{and}\;\; F_c = W_c \setminus K_c 
\end{aligned}
\end{equation}
In TOKI, the word cluster $C_i$ is directly related to the class $c$ if $sim\bigl(\overline{C_{i}^w}, c\bigr) \ge \theta$\textsubscript{\textit{relate}}. Here, $\overline{C_{i}^w}$ is the mean vector of all the embeddings of words in $C_i$, $sim(\cdot, \cdot )$ measures the cosine similarity between two word embeddings, and $\theta$\textsubscript{\textit{relate}} is the threshold relatedness.
After this step, the list of keywords $K_c$ of the class $c$ is identified while the remaining words form a list of non-keywords $F_c$.

While $\theta$\textsubscript{\textit{dist}} needs manual configuration, $\theta$\textsubscript{\textit{relate}} can be automatically estimated by turning it into a binary classification problem. The key idea to determine $\theta$\textsubscript{\textit{relate}} is that related pairs of words can be found via synonyms. To accomplish this, the top 1,000 most common English words are taken from WordNet~\cite{Miller:1995:WordNetALexicalDatabaseforEnglish}. 
Then, TOKI uses Merriam-Webster~\citeN{Dictionary:2002:MerriamWebster}
to find all single-word synonyms of each word, resulting in approximately 32,000 pairs of related words. 
Another 32,000 random pairs of words are generated from WordNet to create a list of unrelated pairs. 
Finally, TOKI determines the value of $\theta$\textsubscript{\textit{relate}} through a binary search on the word embeddings of these lists.
At each iteration, all 64,000 pairs are classified, with each pair of words considered as related if the cosine similarity between two word embeddings is higher than or equal to the current $\theta$\textsubscript{\textit{relate}}. The search stops when precision and recall for both related and unrelated classifications are balanced.
The value of $\theta$\textsubscript{\textit{relate}} varies depending on the word embedding method, as different methods have their unique ways of embedding, thereby impacting the measurement of similarity between words.

\subsection{Trustworthiness Label Computation}
The second pipeline of TOKI, as highlighted in yellow in Figure~\ref{fig:new_approach}, focuses on assessing the trustworthiness of a correct prediction.
The trustworthiness label is determined by comparing the impacts between related and unrelated decision-contributing words based on their total importance scores.
To determine whether a decision-contributing word is related to the class, TOKI uses keywords as anchors to assess semantic relatedness.
We define an indicator function $r(w, c)$ for this purpose by checking whether the nearest word in $W_c$ to $w$ is a keyword, as shown in Equation~\ref{eqn:nearest_word}.
We then formalise the second pipeline in Equation~\ref{eqn:label_compute}.
\begin{equation}\label{eqn:nearest_word}
r(w, c) =
\begin{cases}
1, & \text{if } \underset{\langle w_i, \_ \rangle \in K_c}{\max} sim(w_i, w) \ge \underset{\langle w_i, \_ \rangle \in F_c}{\max} sim(w_i, w), \\
0, & \text{otherwise}.
\end{cases}
\end{equation}
\begin{equation}\label{eqn:label_compute}
\begin{gathered}
    IS\textsubscript{\textit{rel}} = \sum_{\mathclap{\langle w_i, s_i \rangle \in e(f, x_{t})}} s_i*r(w_i, c)
    \,,\;\;
    IS\textsubscript{\textit{unr}} = \sum_{\mathclap{\langle w_i, s_i \rangle \in e(f, x_{t})}} s_i*(1-r(w_i, c))
    \,,\;\text{and}\;\;
    D(f, x_t, c, e) = IS\textsubscript{\textit{rel}} \ge IS\textsubscript{\textit{unr}}
\end{gathered}
\end{equation}
TOKI leverages the explanation method $e(\cdot, \cdot)$ to extract decision-contributing words for the prediction to the given input $x_t$. 
For each word $w$, TOKI identifies the most similar word in the word pool $W_c$ of the predicted class $c$, which is constructed in the first pipeline, by measuring the cosine similarity between their embeddings. 
The semantic relatedness between each word and the class is determined by checking whether the most similar word is a keyword.
Next, TOKI computes the total importance scores $IS$\textsubscript{\textit{rel}} and $IS$\textsubscript{\textit{unr}} for semantically related and unrelated words, respectively. 
Based on the difference between them, the trustworthiness oracle $D$ finally assigns a trustworthiness label (trustworthy or untrustworthy) to the prediction.

Word embeddings themselves can be 
biased due to their training data~\cite{Torregrossa:2021:ASurveyOnTrainingAndEvaluationOfWordEmbeddings}, potentially affecting the ability to measure semantic relatedness.
To mitigate this, TOKI applies ensemble learning~\cite{Leon:2017:EvaluatingTheEffectOfVotingMethodsOnEnsembleBasedClassification} by employing different word embedding methods. 
We use both static embedding methods~\cite{Noam:2016:Swivel:ImprovingEmbeddingsbyNoticingWhatsMissing, Bojanowski:2017:FastText:EnrichingWordVectorswithSubwordInformation}, which produce a single output for each word and contextual embedding methods~\cite{Cer:2018:UniversalSentenceEncoderforEnglish, Devlin::2019:BERTPretrainingofDeepBidirectionalTransformersforLanguageUnderstanding}, which generate different vectors for the same word based on its context.
Each method has a different way of vectorizing words, resulting in different similarity measurements between them.
In the first pipeline, this affects the computation of $\theta$\textsubscript{\textit{relate}} and the keyword identification.
In the second pipeline, different embedding methods identify different similar words in the word pool, leading to different decisions about how the word is related to the class.
Decisions made by all embedding methods are combined using plurality voting~\cite{Leon:2017:EvaluatingTheEffectOfVotingMethodsOnEnsembleBasedClassification} in both pipelines.

\revised{Plurality voting is a simple yet effective voting method where each voter selects a single option, and the option with the most votes wins. In the first pipeline, a word receives a vote from an embedding method if the method identifies it as a keyword. The word is ultimately classified as a keyword if it receives the highest number of votes across all embedding methods. Similarly, in the second pipeline, a word is considered related to a class if it is identified as such by the highest number of embedding methods in the ensemble.}

\section{Targeted Adversarial Attacks on Trustworthiness Vulnerabilities}
\label{sec:attack}

We introduce a novel adversarial attack method guided by TOKI. 
The key idea is \textit{to weaken valid correlations} by replacing words with similar ones that are weakly correlated to the original class, while \textit{strengthening spurious correlations} by injecting unrelated words strongly correlated to other classes.
Table ~\ref{table:compare_trular_a2t} compares 
TOKI-guided attack method with existing adversarial attack methods, exemplified by the SOTA A2T~\cite{Yoo:2021:TowardsImprovingAdversarialTrainingofNLPModels}.

A2T uses the gradient of the loss to determine 
the substitution order of words based on their importance scores in the prediction.
It then iteratively replaces each word with synonyms generated from a counter-fitted word embedding model~\cite{Mrksic:2016:CounterfittingWordVectorstoLinguisticConstraints}.
\revised{This embedding model is injected with antonymy and synonymy constraints into vector space representations to improve its ability to assess semantic similarity.
For example, traditional word embedding models like GloVe~\cite{Pennington:2014:Glove:GlobalVectorsforWordRepresentation} consider ``expensive'' similar to its antonyms, ``cheaper'' and ``inexpensive''. In contrast, the counter-fitted word embedding model prefers synonyms like ``costly'' and ``overpriced''.}
A2T also sets a modification rate to constrain the maximum number of perturbations allowed.
The generated texts are subsequently filtered to ensure part-of-speech consistency and semantic preservation by evaluating the cosine similarity between the sentence encodings of the original and perturbed texts.
A2T has been validated and demonstrated as
a strong adversarial attack method~\cite{Zhou:2024:EvaluatingtheValidityofWordLevelAdversarialAttackswithLargeLanguageModels}.

\begin{table}[t]
    \centering
    \caption{Comparing TOKI-guided attack method (Ours) and A2T~\cite{Yoo:2021:TowardsImprovingAdversarialTrainingofNLPModels} 
    }
    \resizebox{0.83\textwidth}{!}{
    \begin{tabular}{|l|c|c|}
        \hline
        \makecell[c]{\textbf{Components}} & \multicolumn{1}{l|}{\makecell[c]{\textbf{TOKI-guided attack method (Ours)}}} & \multicolumn{1}{l|}{\makecell[c]{\textbf{A2T}~\cite{Yoo:2021:TowardsImprovingAdversarialTrainingofNLPModels}}} \\
        \hline
        Word Ranking Method &  Gradient-based Word Importance & Gradient-based Word Importance \\
        \hline
        Source of Synonyms & \textit{Trustworthiness Oracle} & {Counter-fitted Embedding} \\
        \hline
        Word Substitution & Word Embedding + \textit{Word-Class Correlation} & Word Embedding \\
        \hline
        Constraints & \makecell{Modification Rate\\DistilBERT Cosine Similarity\\Part-of-Speech Consistency} & \makecell{Modification Rate\\DistilBERT Cosine Similarity\\Part-of-Speech Consistency} \\
        \hline
    \end{tabular}%
    }
    \label{table:compare_trular_a2t}
\end{table}
\begin{figure}[t]
    \centering
    \includegraphics[width=0.92\linewidth, clip, trim=0 29mm 0 0]{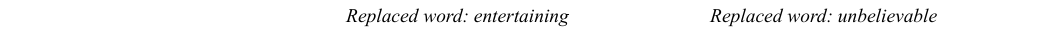}
    \includegraphics[width=0.92\linewidth, clip, trim=0 0 0 6.5mm]{figures/trular_a2t_example.pdf}
    \vspace{-1mm}
    \caption{Adversarial examples generated by TOKI (Ours) and A2T. 
    Words highlighted in \textcolor{myorange}{orange} or \textcolor{myblue}{blue} show their contribution to the \textcolor{myorange}{negative} or \textcolor{myblue}{positive} class, with the shade indicating their level of importance.
    }
    \label{fig:an_attack_example}
    \Description[]{}
\end{figure}

Our method follows the same architecture as A2T. The key difference is that it finds synonyms in word pools $W$ constructed by TOKI, based on word-class correlations measured by the averaged importance scores $\overline{s}$.
To replace a word, TOKI-guided attack method checks whether the word is related to the original predicted class.
If it is, the word is replaced with a similar keyword of that class that has a \textit{low importance score}.
Otherwise, it is replaced with a similar non-keyword from other classes that has a \textit{high importance score}.
This mechanism tricks the model into using unrelated words as cues for predictions to cause misclassification,
increasing the likelihood of successful attacks while reducing the number of perturbations.

Figure~\ref{fig:an_attack_example} compares adversarial examples generated by TOKI and A2T on the same input, showing that A2T requires more perturbations to succeed.
The text is initially predicted as positive mainly based on the words ``entertaining'' and ``unbelievable''.
A2T first substitutes ``entertaining'' with ``fun'', a synonym identified by the counter-fitted word embedding model.
However, this change fails to attack the model, as the positive clues ``fun'' and ``unbelievable'' still dominate.
A2T then replaces ``unbelievable'' with its synonym ``extraordinary''.
This time, the overall impact of positive clues is reduced, allowing the negative clue ``awfully'' to dominate.
A2T substitutes words based only on the semantic meaning, requiring two substitutions to create a successful adversarial example.
In contrast, TOKI-guided attack method replaces ``entertaining'' with a similar word ``humorous''.
This word is deemed weakly correlated to the positive class by TOKI in this case, allowing the negative clues to dominate and alter the prediction with just one substitution.

\section{Trustworthiness Benchmark}
\label{sec:benchmark}
Trustworthiness oracles aim to determine whether a prediction is trustworthy, as described in Section~\ref{sec:problem_definition}. 
Evaluating them requires datasets containing model predictions, corresponding explanations, and trustworthiness labels (trustworthy or untrustworthy). 
We call them ``trustworthiness datasets'' to distinguish them from ``datasets'' used to train and validate ML models.
However, limited trustworthiness datasets are available for such evaluation~\cite{Schlegel:2022:TowardsHumanCentredExplainabilityBenchmarksForTextClassification}. 

\textit{\textbf{Dong's trustworthiness dataset}.}
The human-based evaluation conducted by \citeN{Dong:2018:ComparingAutomaticandHumanEvaluationofLocalExplanationsforTextClassification} is one of the most relevant studies. 
This evaluation uses two datasets: a subset~\cite{Ribeiro:2016:WhyShouldITrustYouExplainingthePredictionsofAnyClassifier} of the 20 newsgroups (\textit{20news}) differentiating Christianity from Atheism, and \textit{movie} reviews with sentiment labels.
In the evaluation, text inputs, highlighting the top words identified by the explanation method, are shown to crowdworkers. 
They then guess the system's output and state their confidence on a five-point Likert scale, ranging from ``strongly disagree'' to ``strongly agree''.
To derive a trustworthiness dataset,
only answers, where the model correctly predicts the output, are selected from the original crowdworkers' responses.
The answers, where crowdworkers either guess incorrectly with high confidence (4--5) or correctly with low confidence (1--2), are deemed untrustworthy.
Other answers, where the crowdworkers guess correctly with high confidence (4--5), are trustworthy.
Trustworthiness labels from the crowdworkers for the same prediction are combined using plurality voting~\cite{Leon:2017:EvaluatingTheEffectOfVotingMethodsOnEnsembleBasedClassification} to determine the final trustworthiness label. 


\begin{table}[t]
    \caption{Trustworthiness benchmark 
    }
    \label{table:ground_truths}
    \centering
    \resizebox{\textwidth}{!}{
    \begin{tabular}{|l|l|c|c|c|c|c|c|c|}
    \hline
    \multicolumn{2}{|c|}{\multirow{2}{*}{\textbf{Dataset}}} & \multicolumn{3}{c|}{\textbf{Data Statistics}} & \multicolumn{2}{c|}{\textbf{Model Under Test}} &  \multirow{2}{*}{\makecell{\textbf{Number of}\\\textbf{top words}}} &  \multirow{2}{*}{\makecell{\textbf{Importance}\\\textbf{Type}}} \\
    \cline{3-7}
    \multicolumn{2}{|c|}{} & \multicolumn{1}{p{1.15cm}|}{\makecell[c]{\small \textbf{Trust}}} & \multicolumn{1}{p{1.15cm}|}{\makecell[c]{\small \textbf{Untrust}}} & \multicolumn{1}{p{1.15cm}|}{\makecell[c]{\small \textbf{Total}}} & \multicolumn{1}{c|}{\makecell[c]{\small \textbf{Model Type}}} & \small \textbf{Accuracy} &  & \\
    \hline
    \multicolumn{2}{|l|}{movie~\cite{Dong:2018:ComparingAutomaticandHumanEvaluationofLocalExplanationsforTextClassification}}
    & \multirow{2}{*}{311}     & \multirow{2}{*}{47}        & \multirow{2}{*}{358} & \multirow{2}{*}{\makecell{Multilayer perceptron\\(MLPs)}} & 0.832 & \multirow{2}{*}{10, 20} & importance \\
    \multicolumn{2}{|l|}{20news~\cite{Dong:2018:ComparingAutomaticandHumanEvaluationofLocalExplanationsforTextClassification}}
     &       &          &   &  & 0.939 &   & equivalent \\
    \hline
    \multicolumn{2}{|l|}{CAMS~\cite{Garg:2022:CAMS:AnAnnotatedCorpusforCausalAnalysisofMentalHealthIssuesinSocialMediaPosts}}
    & 1,206 & 739 & 1,945 & 
    \makecell{
    \href{https://huggingface.co/Tianlin668/CAMS}{mentalbert-base-}\\\href{https://huggingface.co/Tianlin668/CAMS}{uncased}} & 0.397 & 10 & \makecell{importance\\equivalent} \\
    \hline
    \multicolumn{2}{|l|}{\makecell[l]{HateXplain\\\cite{Mathew:2021:HateXplain:ABenchmarkDatasetforExplainableHateSpeechDetection}}}
    & 3,002 & 304 & 3,306 & \href{https://huggingface.co/Hate-speech-CNERG/bert-base-uncased-hatexplain-rationale-two}{bert-base-uncased} & 0.797 & 10 & \makecell{importance\\equivalent} \\
    \hline
    \multicolumn{2}{|l|}{\revised{Issues~\cite{Schulte:2024:StudyingTheExplanationsForTheAutomatedPredictionOfBugAndNonbugIssuesUsingLimeAndShap}}}
    & \revised{2,187} & \revised{45} & \revised{2,232} & \makecell[c]{\revised{sebert-base}
    } 
    & \revised{0.945} & \revised{10} & \makecell{\revised{importance}\\\revised{different}} \\
    \hline
    \multirow{6}{*}{\makecell[l]{Ours}} & amazon\_polarity            & \multirow{6}{*}{226}      & \multirow{6}{*}{19}         & \multirow{6}{*}{245} & \href{https://huggingface.co/pig4431/amazonPolarity\_roBERTa\_5E}{roberta-base-cased}  & 0.960 & \multirow{6}{*}{5, 10, 20} & \multirow{6}{*}{\makecell{{importance}\\{different}}}  \\
     & ag\_news          &        &           &    & \href{https://huggingface.co/mrm8488/bert-mini-finetuned-age\_news-classification}{bert-base-uncased} & 0.934 & &   \\
     & rotten\_tomatoes  &        &           &   &  \href{https://huggingface.co/xianzhew/distilbert-base-uncased\_rotten\_tomatoes}{distilbert-base-uncased}   & 0.841 & &  \\
     & yahoo\_answers\_topics    &        &           &   & \href{https://huggingface.co/fabriceyhc/bert-base-uncased-yahoo\_answers\_topics}{bert-base-uncased}       & 0.750 & & \\
     & imdb              &        &           &   & \href{https://huggingface.co/lvwerra/distilbert-imdb}{distilbert-base-uncased}                                     & 0.928  & & \\
     & emotion           &        &           &    & \href{https://huggingface.co/sabre-code/distilbert-base-uncased-finetuned-emotion}{distilbert-base-uncased} & 0.926 & & \\
    \hline
    \end{tabular}%
    }
\end{table}

\textit{\textbf{CAMS and HateXplain}.}
CAMS~\cite{Garg:2022:CAMS:AnAnnotatedCorpusforCausalAnalysisofMentalHealthIssuesinSocialMediaPosts} is a corpus for classifying mental health issues from social media posts, while HateXplain~\cite{Mathew:2021:HateXplain:ABenchmarkDatasetforExplainableHateSpeechDetection} is a dataset for hate speech detection. 
Both datasets provide ground-truth explanations for each instance. 
In CAMS, annotations highlight phrases used as inferences for predictions.
In HateXplain, 
tokens are labeled as 0 or 1 to indicate whether they are part of the explanation.
Trustworthiness datasets are created by explaining the models and comparing these explanations with the ground-truth ones.
To measure the plausibility of the reasoning behind predictions, we follow \citeN{ElZini:2022:OntheEvaluationofthePlausibilityandFaithfulnessofSentimentAnalysisExplanations} and use \textit{explanation precision} \( = |E~\cap~G|~/~|E|\), where $E$ is the model explanation and $G$ is the ground-truth explanation.
High precision suggests that the model explanation is unlikely to provide a word not in the ground-truth explanation.
A prediction is then considered trustworthy if its explanation precision $\ge$ 0.5.


\textit{\revised{\textbf{Issues: A SE-specific dataset}}.} 
\revised{We adopt the study of \citeN{Schulte:2024:StudyingTheExplanationsForTheAutomatedPredictionOfBugAndNonbugIssuesUsingLimeAndShap}, which analyses explanations for the automated classification of bug and non-bug issues, a critical task in SE.
These issues are reported in issue tracking systems, such as JIRA or Github.
Each prediction is explained by explanation methods, such as LIME~\cite{Ribeiro:2016:WhyShouldITrustYouExplainingthePredictionsofAnyClassifier}. 
The authors then review the explanation based on multiple criteria, 
assigning
a score of +1 if 
satisfied, 0 if 
neutral, and -1 if 
not. 
To assess a prediction's trustworthiness, we focus on two criteria: \textit{related} and \textit{unambiguous}.
Specifically, \textit{related} means there is a clear relationship between the important words of the explanation and the prediction.
On the other hand, \textit{unambiguous} implies there are no words of mixed meaning or all words are used in their correct meaning with respect to the explanation.
Predictions are 
deemed trustworthy 
if the average score of these two criteria is greater than 0.
If the average score is lower than 0, they are deemed untrustworthy.
We finally combine these decisions of all annotators using plurality voting~\cite{Leon:2017:EvaluatingTheEffectOfVotingMethodsOnEnsembleBasedClassification} to determine the final trustworthiness label for each prediction.}

\textit{\textbf{Importance-different trustworthiness dataset}.}
We create an additional trustworthiness dataset that considers the distribution of importance scores of words.
Six datasets and corresponding models are selected from Huggingface: \textit{amazon\_polarity},
\textit{ag\_news},
\textit{rotten\_tomatoes},
\textit{yahoo\_answers\_topics},
\textit{imdb},
and \textit{emotion}.
We randomly sample 1,000 predictions and have three trained participants annotating their trustworthiness.
For each prediction, the annotators first guess the output of the text input.
They then review the model prediction and its explanation generated by LIME~\cite{Ribeiro:2016:WhyShouldITrustYouExplainingthePredictionsofAnyClassifier}, a SOTA method known for its effectiveness and faithfulness~\cite{Mariotti:2024:TextFocus:AssessingtheFaithfulnessofFeatureAttributionMethodsExplanationsinNaturalLanguageProcessing, Zhao:2024:ExplainabilityforLargeLanguageModelsASurvey}.
Finally, the prediction is manually labelled as trustworthy or untrustworthy by annotators. 
Only predictions where both the model and annotators guess the correct output are considered.
The trustworthiness label for each prediction is determined by plurality voting~\cite{Leon:2017:EvaluatingTheEffectOfVotingMethodsOnEnsembleBasedClassification} based on all annotations.
It is observed that the annotators often deem a prediction untrustworthy if unrelated words receive significantly higher importance scores in its explanation. 
\textit{\textbf{Explanation method}.}
Three methods are used to extract decision-contributing words.
\begin{itemize}
    \item \textit{LIME}
    approximates the model locally with an interpretable model on perturbed samples created around the input~\cite{Ribeiro:2016:WhyShouldITrustYouExplainingthePredictionsofAnyClassifier}. 
    Our experiments use 5,000 perturbed samples.
    \item \textit{Word omission}~\cite{Li:2017:UnderstandingNeuralNetworksthroughRepresentationErasure} estimates the contribution of individual words by deleting them and measuring the change in probability for the predicted class~\cite{Dong:2018:ComparingAutomaticandHumanEvaluationofLocalExplanationsforTextClassification}.
    \item \textit{Gradient} computes the output gradient 
    with respect to the input~\cite{Mohebbi:2021:ExploringtheRoleofBERTTokenRepresentationstoExplainSentenceProbingResults}.
\end{itemize}

\textit{\textbf{Models under test}.}
For our new datasets, popular fine-tuned models from Huggingface are chosen. 
For the remaining datasets, models provided by the authors
are used.
Table~\ref{table:ground_truths} summarises the final trustworthiness benchmark, the models under test, and their corresponding accuracies.

\section{Evaluation}
\label{sec:evaluations}
This section describes a series of experiments conducted to evaluate how well TOKI addresses the trustworthiness oracle problem and adversarial attacks text classifiers.

\subsection{Baselines} \label{sec:expr_baselines}

We adopt a naive approach, named \textbf{Naive}, which assesses trustworthiness based on model confidence.
Naive considers a prediction untrustworthy if its confidence is lower than a threshold $\theta$\textsubscript{\textit{conf}}. 
For ablation studies, we use a TOKI's variant, called \textbf{TOKI (–K.I)}, which directly measures the relatedness between decision-contributing words and class names without identifying keywords~\cite{Cho:2024:TOWER}.
We also compare TOKI-guided attack method with \textbf{A2T}~\cite{Yoo:2021:TowardsImprovingAdversarialTrainingofNLPModels}, a SOTA adversarial attack method that has been validated to be strong and effective~\cite{Zhou:2024:EvaluatingtheValidityofWordLevelAdversarialAttackswithLargeLanguageModels}.



\subsection{Research Questions} \label{sec:research_questions}

We answer three research questions to investigate the trustworthiness problem and evaluate TOKI.
\begin{enumerate}[label=\textbf{RQ\arabic*.}, leftmargin=3.4\parindent]
    \item Does a prediction's uncertainty reflect its trustworthiness?
    \begin{enumerate}[label*=\textbf{\arabic*.}, leftmargin=2\parindent] 
        \item 
        How are prediction uncertainty and trustworthiness related?
        \item What is the optimal configuration for $\theta$\textsubscript{\textit{conf}} of Naive? 
    \end{enumerate}
    

    \item How effective and efficient is TOKI in trustworthiness assessment?
    \begin{enumerate}[label*=\textbf{\arabic*.}, leftmargin=2\parindent] 
        \item What is the optimal configuration for $\theta$\textsubscript{\textit{dist}}?
        \item How effective is TOKI compared to Naive?
        \item How does identifying keywords affect TOKI's effectiveness and efficiency?
        \item What is the impact of different explanation methods on TOKI?
    \end{enumerate}
        
    \item How effective is TOKI-guided attack method compared to A2T?
\end{enumerate}

\subsection{Metrics} \label{sec:metrics} 


\subsubsection*{\revised{Trustworthiness assessment}}
\revised{We emphasise that our evaluations focus on the effectiveness of trustworthiness oracles in classifying predictions as trustworthy or not, rather than assessing the trustworthiness of ML models themselves. 
The effectiveness of trustworthiness oracles is determined by the alignment between their generated trustworthiness labels and human-annotated ones that reflect human perceptions of trustworthiness in the benchmark shown in Table~\ref{table:ground_truths}. We frame trustworthiness oracles as binary classifiers with trustworthy and untrustworthy labels and 
evaluate their effectiveness using standard performance metrics for binary classification.
\begin{itemize}
    \item \textit{Accuracy}: the proportion of predictions correctly labelled as trustworthy or untrustworthy.
    \item \textit{Precision}, \textit{sensitivity}, and \textit{F1-score}: the performance in detecting trustworthy predictions.
    \item \textit{Specificity}: the performance in detecting untrustworthy predictions.
    \item \textit{Geometric mean} (\textit{G-mean)}:
    the balance between the classification performance on trustworthy and untrustworthy predictions, computed as \(\sqrt{\text{\textit{sensitivity}}  \times \text{\textit{specificity}}}\).
\end{itemize}
This usage of these metrics differs from that of model confidence in \textbf{RQ1}, which is assumed to be an unsuitable indicator of trustworthiness. 
G-mean serves as a balanced metric for evaluating classification performance on both trustworthy and untrustworthy predictions.
A higher G-mean indicates better alignment between the trustworthiness benchmark and the oracles.
Additionally, the efficiency of the trustworthiness oracles is assessed by their processing time in seconds.}

\subsubsection*{Adversarial attack}
We report the attack success rate (\textit{\textbf{ASR}}), defined as $\frac{\text{\textit{\# of successful attacks}}}{\text{\textit{\# of total attacks}}}$, and the number of perturbations (\textbf{\textit{NP}}) to evaluate effectiveness.
We also use \textbf{\textit{Bert}}~\cite{Devlin::2019:BERTPretrainingofDeepBidirectionalTransformersforLanguageUnderstanding}
and \textbf{\textit{USE}}~\cite{Cer:2018:UniversalSentenceEncoderforEnglish} scores to assess cosine similarity between original and adversarial examples.

\subsection{Experimental Setup} \label{sec:experimental_setup}

We use six word embedding methods for TOKI's word embedding ensemble model.
    \ding{172}~\textbf{NNLM}~\cite{Bengio:2000:ANeuralProbabilisticLanguageModel} learns embeddings 
    and language models 
    using a feedforward neural network. 
    \ding{173}~\textbf{GloVe}~\cite{Pennington:2014:Glove:GlobalVectorsforWordRepresentation} generates word embeddings from corpus word-to-word co-occurrence matrices. 
    \ding{174}~\textbf{Swivel}~\cite{Noam:2016:Swivel:ImprovingEmbeddingsbyNoticingWhatsMissing} generates low-dimensional embeddings from feature co-occurrence matrices. 
    \ding{175}~\textbf{FastText}~\cite{Bojanowski:2017:FastText:EnrichingWordVectorswithSubwordInformation} represents each word as a bag of character n-grams,
    capable of embedding misspelled, rare, or out-of-vocabulary words.
    \ding{176}~\textbf{USE}~\cite{Cer:2018:UniversalSentenceEncoderforEnglish} encodes sentences into embeddings for transfer learning to other tasks.
    \ding{177}~\textbf{Bert}~\cite{Devlin::2019:BERTPretrainingofDeepBidirectionalTransformersforLanguageUnderstanding} is the first deeply bidirectional, unsupervised language representation, pre-trained on a large text corpus to condition both left and right contexts of each word.


We use the trustworthiness benchmark described in Section~\ref{sec:benchmark} to evaluate trustworthiness oracles. 
For our datasets, 2,000 training instances are randomly sampled for explanations in TOKI's first pipeline, while all training instances are used in the remaining datasets.
Trustworthiness oracles employing LIME, omission, and gradient explanations are denoted by the suffixes ``-lime'', ``-omis'', and ``-grad'', respectively.
Regarding adversarial attacks, we implement our method using TextAttack~\cite{Morris:2020:TextAttack:AFrameworkforAdversarialAttacksDataAugmentationandAdversarialTraininginNLP}, which already includes A2T's implementation~\cite{Yoo:2021:TowardsImprovingAdversarialTrainingofNLPModels}, and use A2T's default settings for both.
We then attack random samples of up to 8,000 instances from each dataset in Table~\ref{table:ground_truths}.
Experiments are run 10 times on a Macbook M3 Pro with 12-core CPU, 18-core GPU, 18GB RAM, and 512GB SSD.
The final results are averaged across these runs.


\subsection{Results} \label{sec:experimental_results}
\subsubsection*{\textbf{RQ1}: Relation between prediction uncertainty and trustworthiness}
\begin{figure}[t]
    \centering
    \begin{subfigure}[ht]{0.195\linewidth}
        \centering
        \includegraphics[width=1.0\linewidth, clip, trim=1mm 2mm 1mm 2mm]{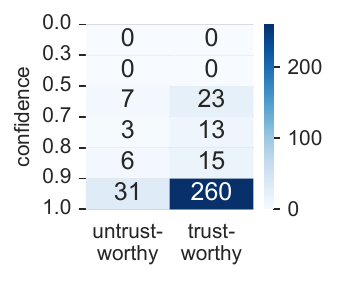}
        \vspace{-5mm}
        \caption{Dong's}
    \end{subfigure}%
    \hfill
    \begin{subfigure}[ht]{0.195\linewidth}
        \centering
        \includegraphics[width=1.0\linewidth, clip, trim=1mm 2mm 1mm 2mm]{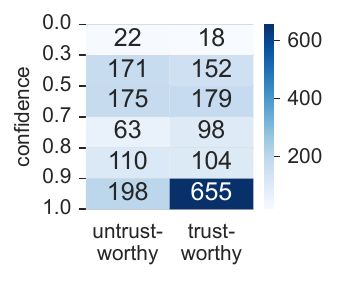}
        \vspace{-5mm}
        \caption{CAMS}
    \end{subfigure}
    \hfill
    \begin{subfigure}[ht]{0.195\linewidth}
        \centering
        \includegraphics[width=1.0\linewidth, clip, trim=1mm 2mm 1mm 2mm]{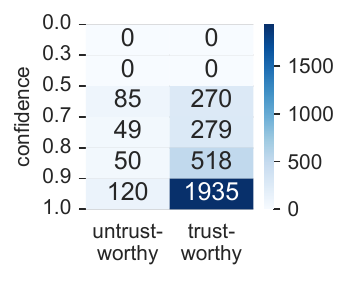}
        \vspace{-5mm}
        \caption{HateXplain}
    \end{subfigure}
    \hfill
    \begin{subfigure}[ht]{0.195\linewidth}
        \centering
        \includegraphics[width=1.0\linewidth, clip, trim=1mm 2mm 1mm 2mm]{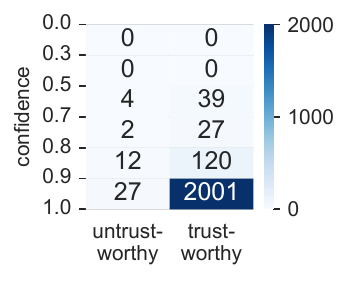}
        \vspace{-5mm}
        \caption{\revised{Issues}}
    \end{subfigure}
    \hfill
    \begin{subfigure}[ht]{0.195\linewidth}
        \centering
        \includegraphics[width=1.0\linewidth, clip, trim=1mm 2mm 1mm 2mm]{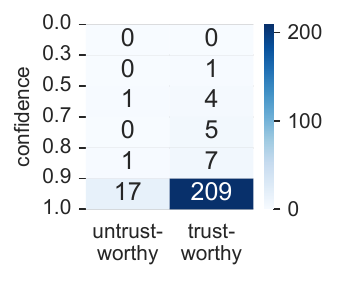}
        \vspace{-5mm}
        \caption{Ours}
    \end{subfigure}
    \vspace{-2mm}
    \caption{{The distribution of predictions' confidence across trustworthiness labels.}}
    \label{fig:correlation_trust_conf}
    \Description[Highly confident predictions are more likely to be trustworthy. However, strong confidence predictions can also be untrustworthy, and those with low confidence can still be trustworthy]{Fully described in the text.}    
\end{figure}

We use model confidence as a metric to assess the uncertainty of ML predictions.
Figure~\ref{fig:correlation_trust_conf} addresses \textbf{RQ1.1} by illustrating the distribution of model confidence for two trustworthiness labels on the benchmark shown in Table~\ref{table:ground_truths}.
The majority of high confidence predictions are trustworthy. 
Highly confident (0.9--1.0) predictions that are trustworthy account for 76\%--99\% of predictions with highly strong confidence.
Similarly, trustworthy predictions with high confidence (0.8--0.9) represent 49\%--91\% of high confidence predictions.
While predictions with strong confidence are likely to be trustworthy, several confident predictions are still untrustworthy.
Confident (0.8--1.0) but untrustworthy predictions make up 2\%--29\% of all confident predictions.
In contrast, predictions with low confidence ($<$0.8) can also be trustworthy, 
with 51\%--92\% of low confidence predictions being trustworthy.

\begin{figure}
    \centering
    \begin{minipage}{\textwidth}
    \centering
        \includegraphics[width=0.66\textwidth, clip, trim=0mm 0mm 0cm 0mm]{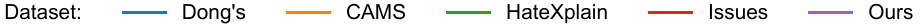} 
        \Description[Legend]{Fully described in the text.}
    \end{minipage}
    
    \begin{minipage}{0.325\textwidth}
        \centering
        \includegraphics[width=\textwidth, clip, trim=2mm 2mm 0 -1.5mm]{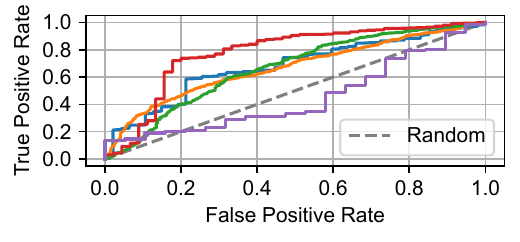} 
        \vspace{-6mm}
        \caption{\revised{ROC curves of Naive.}}
        \Description[ROC curve of Naive.]{Fully described in the text.}
        \label{fig:roc_tconf}
    \end{minipage}
    \hfill
    \begin{minipage}{0.325\textwidth}
        \centering
        \includegraphics[width=\textwidth, clip, trim=2mm 2mm 0 0mm]{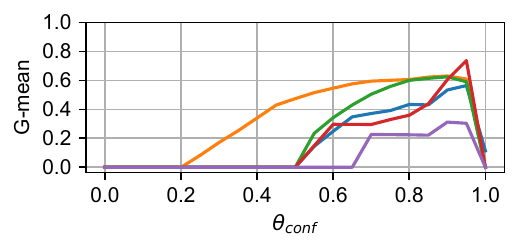} 
        \vspace{-6mm}
        \caption{\revised{Naive with different $\theta$\textsubscript{\textit{conf}}.}}
        \label{fig:thres_conf_chart}
        \Description[Threshold of Naive.]{Fully described in the text.}
    \end{minipage}
    \hfill
    \begin{minipage}{0.325\textwidth}
        \centering
        \includegraphics[width=\textwidth, clip, trim=2mm 2mm 0 0mm]{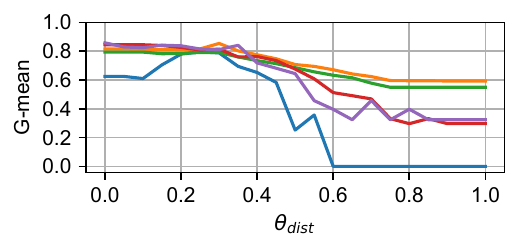} 
        \vspace{-6mm}
        \caption{\revised{TOKI with different $\theta$\textsubscript{\textit{dist}}.}}
        \label{fig:experimental_result_rq1}
        \Description[Threshold of TOKI.]{Fully described in the text.}
    \end{minipage}
\end{figure}

We now address \textbf{RQ1.2}, finding the optimal value of $\theta$\textsubscript{\textit{conf}} for Naive, which relies solely on model confidence. 
Figure~\ref{fig:roc_tconf} shows receiver-operating characteristic (ROC) curves of Naive across trustworthiness datasets.
\revised{Only the ROC curve for \textit{Issues} is slightly closer to the top-left corner of the plot, while the others are near the random curve. Notably, the ROC curve for the new trustworthiness dataset falls below the random curve.
This indicates that model confidence is limited in distinguishing between trustworthy and untrustworthy predictions.}
\revised{Figure~\ref{fig:correlation_trust_conf} further shows that increasing $\theta$\textsubscript{\textit{conf}} on the y-axis decreases the false positive rate, where untrustworthy predictions are labeled as trustworthy, but also increases the false negative rate, where trustworthy predictions are labeled as untrustworthy.
In other words, increasing $\theta$\textsubscript{\textit{conf}} enhances specificity but reduces sensitivity.
Hence, we choose to maximise G-mean as the criterion to find $\theta$\textsubscript{\textit{conf}}.}
Figure~\ref{fig:thres_conf_chart} illustrates Naive's effectiveness in G-mean with different values of $\theta$\textsubscript{\textit{conf}}.
As the value of 0.9 achieves the best G-mean for Naive, we set $\theta$\textsubscript{\textit{conf}} to 0.9 for the remaining experiments.


\begin{custombox}[Answer to RQ1]
Uncertainty metrics of ML predictions, such as model confidence, are not suitable for assessing their trustworthiness.
Although high confidence predictions tend to be trustworthy, relying on these metrics easily overlooks both trustworthy low confidence predictions and untrustworthy high confidence predictions.
\end{custombox}

\subsubsection*{\textbf{RQ2}: Effectiveness and Efficiency of TOKI}


\revised{Intuitively, increasing $\theta$\textsubscript{\textit{dist}} leads to an over-inclusion of words as keywords, which results in a bias towards classifying predictions as trustworthy. 
This increases sensitivity while reducing specificity. 
Conversely, decreasing $\theta$\textsubscript{\textit{dist}} causes related keywords to be overlooked, leading to a bias towards classifications of predictions as untrustworthy, thereby reducing sensitivity and increasing specificity.
Hence, we also address \textbf{RQ2.1} by measuring the effectiveness of TOKI-lime with various values of $\theta$\textsubscript{\textit{dist}} using G-mean, a metric that balances sensitivity and specificity.}
Figure~\ref{fig:experimental_result_rq1} reveals that the optimal value of $\theta$\textsubscript{\textit{dist}} ranges from 0.2 to 0.4, depending on the models under test. 
Across all trustworthiness datasets, when $\theta$\textsubscript{\textit{dist}} exceeds 0.6,  G-mean drops significantly. 
In this case, TOKI considers all words as keywords and biases for labeling predictions as trustworthy.
In the subsequent experiments, we set $\theta$\textsubscript{\textit{dist}} to 0.3, as it balances the effectiveness across the benchmark.

\begin{table*}[t]
    \caption{Comparison between TOKI (Ours), Naive, and a TOKI's variant that is without keyword identification (indicated by the suffix --K.I) against the trustworthiness benchmark}
    \label{table:experimental_result_rq2_5_6}
    \resizebox{\textwidth}{!}{
    \begin{tabular}{|c|l|C|C|C|C|C|C|c|c|c|c|}
        \hline
        \textbf{Dataset} & \multicolumn{1}{p{2.6cm}|}{\makecell[c]{\textbf{Method}}}     & \multicolumn{1}{@{}p{1.4cm}@{}|}{\makecell[c]{\textbf{Acc} ($\uparrow$)}}   & \multicolumn{1}{@{}p{1.4cm}@{}|}{\makecell[c]{\textbf{Pre} ($\uparrow$)}}   & \multicolumn{1}{@{}p{1.4cm}@{}|}{\makecell[c]{\textbf{Sen} ($\uparrow$)}}   & \multicolumn{1}{@{}p{1.4cm}@{}|}{\makecell[c]{\textbf{F1} ($\uparrow$)}} & \multicolumn{1}{@{}p{1.4cm}@{}|}{\makecell[c]{\textbf{Spec} ($\uparrow$)}}  & \multicolumn{1}{@{}p{1.4cm}@{}|}{\makecell[c]{\textbf{G-mean}\\($\uparrow$)}} & \makecell{\small \textbf{Training}\\\small \textbf{Instances}} & \makecell{\footnotesize \textbf{Keyword}\\\footnotesize \textbf{Identification}\\\footnotesize \textbf{Time} ($\downarrow$s)} & \makecell{\small \textbf{Test}\\\small \textbf{Instances}} & \makecell{\footnotesize \textbf{Trustworthiness}\\\footnotesize \textbf{Label} \textbf{Computa-}\\\footnotesize \textbf{tion} \textbf{Time} ($\downarrow$s)} \\
        \thickhline
        \multirow{7}{*}{\rotatebox[origin=c]{90}{\makecell{Dong's\\\cite{Dong:2018:ComparingAutomaticandHumanEvaluationofLocalExplanationsforTextClassification}}}} & TOKI-lime \small (Ours) \normalsize  & 0.930 & 0.947 & 0.974 & 0.960    & 0.638 & 0.789  & \multirow{3}{*}{2,509} & 4,866                        & \multirow{7}{*}{358} & 1,053                \\
        & TOKI-omis \small (Ours) \normalsize & 0.882 & 0.912 & 0.956 & 0.934    & 0.392 & 0.612  &                       & 98                          &                      & 379                \\
        & TOKI-grad \small (Ours) \normalsize  & 0.853 & 0.944 & 0.891 & 0.917    & 0.387 & 0.587  &                       & 41                          &                      & 704                \\
        \cline{2-10}
        \cline{12-12}
        & Naive                      & 0.771 & 0.893 & 0.836 & 0.864    & 0.340 & 0.533  & \xmark                & \xmark                      &                      & \xmark             \\
        \cline{2-10}
        \cline{12-12}
        & TOKI-lime \small (--K.I) \normalsize            & 0.302 & 0.985 & 0.209 & 0.345    & 0.915 & 0.437  & \multirow{3}{*}{\xmark}                & \multirow{3}{*}{\xmark}                      &                      & 1,051                \\
        & TOKI-omis \small (--K.I) \normalsize                & 0.285 & 0.956 & 0.192 & 0.320    & 0.902 & 0.416  &                 &                       &                      & 377                \\
        & TOKI-grad \small (--K.I) \normalsize               & 0.233 & 1.000 & 0.170 & 0.290    & 0.968 & 0.406  &                 &                       &                      & 702                \\
        \thickhline
        \multirow{7}{*}{\rotatebox[origin=c]{90}{\makecell{CAMS\\\cite{Garg:2022:CAMS:AnAnnotatedCorpusforCausalAnalysisofMentalHealthIssuesinSocialMediaPosts}}}} & TOKI-lime  \small (Ours) \normalsize    & 0.878 & 0.872 & 0.941 & 0.905    & 0.775 & 0.854  & \multirow{3}{*}{2,158} & 198,359                        & \multirow{7}{*}{1,945} & 224,540                \\
        & TOKI-omis \small (Ours) \normalsize  & 0.735 & 0.840 & 0.622 & 0.715    & 0.864 & 0.733  &                       & 30,030                          &                      & 31,968                \\
        & TOKI-grad \small (Ours) \normalsize  & 0.539 & 0.888 & 0.499 & 0.639    & 0.719 & 0.599  &                       & 2,366                          &                      & 3,116                \\
        \cline{2-10}
        \cline{12-12}
         & Naive                      & 0.615 & 0.768 & 0.543 & 0.636    & 0.732 & 0.631  & \xmark                & \xmark                      &                      & \xmark             \\
        \cline{2-10}
        \cline{12-12}
         & TOKI-lime \small (--K.I) \normalsize                      & 0.422 & 0.648 & 0.343 & 0.449    & 0.551 & 0.435  & \multirow{3}{*}{\xmark}                & \multirow{3}{*}{\xmark}                      &                      & 224,435             \\
        & TOKI-omis \small (--K.I) \normalsize  & 0.442 & 0.557 & 0.232 & 0.327    & 0.683 & 0.398  &                       &                           &                      & 31,912                \\
        & TOKI-grad \small (--K.I) \normalsize  & 0.388 & 0.836 & 0.356 & 0.500    & 0.531 & 0.435  &                       &                           &                      & 3,071                \\
        \thickhline
        \multirow{7}{*}{\rotatebox[origin=c]{90}{\parbox[c]{1.6cm}{\centering HateXplain\\\cite{Mathew:2021:HateXplain:ABenchmarkDatasetforExplainableHateSpeechDetection}}}} & TOKI-lime  \small (Ours) \normalsize    & 0.933 & 0.966 & 0.961 & 0.963    & 0.661 & 0.797  & \multirow{3}{*}{2,364} & 352,721                        & \multirow{7}{*}{3,306} &  464,720               \\
        & TOKI-omis \small (Ours) \normalsize & 0.855 & 0.928 & 0.904 & 0.916    & 0.538 & 0.697  &                       & 4,630                          &                      & 4,913                \\
        & TOKI-grad \small (Ours) \normalsize & 0.684 & 0.375 & 0.385 & 0.380    & 0.785 & 0.550  &                       & 5,334                          &                      & 7,303                \\
        \cline{2-10}
        \cline{12-12}
         & Naive                      & 0.640  & 0.942  & 0.645  &  0.765    & 0.605  & 0.624   & \xmark                & \xmark                      &                      & \xmark             \\
        \cline{2-10}
        \cline{12-12}
         & TOKI-lime \small (--K.I) \normalsize      & 0.103   & 0.953  & 0.014  & 0.027     & 0.984  & 0.117  & \multirow{3}{*}{\xmark}                & \multirow{3}{*}{\xmark}                       &                      & 464,606             \\
        & TOKI-omis \small (--K.I) \normalsize  & 0.137 & 0.879 & 0.010 & 0.020    & 0.975 & 0.099  &                       &                           &                      & 4,818                \\
        & TOKI-grad \small (--K.I) \normalsize  & 0.737 & 0.586 & 0.020 & 0.039    & 0.979 & 0.140  &                       &                           &                      & 7,271                \\
        \thickhline
        \multirow{3}{*}{\rotatebox[origin=c]{90}{\parbox[c]{1.2cm}{\linespread{0.7}\selectfont\centering {Issues\\\cite{Schulte:2024:StudyingTheExplanationsForTheAutomatedPredictionOfBugAndNonbugIssuesUsingLimeAndShap}}}}} & TOKI-lime \small (Ours) \normalsize   & 0.958 & 0.993 & 0.963 & 0.978    & 0.689 & 0.815  & 3,090                  & 585,523                        & \multirow{3}{*}{2,232}  &  353,754  \\
        \cline{2-10}
        \cline{12-12}
        & Naive                      & 0.944 & 0.983 & 0.959 & 0.971    & 0.200 & 0.312  & \xmark                & \xmark                      &                      & \xmark             \\
        \cline{2-10}
        \cline{12-12}
        & TOKI-lime \small (--K.I) \normalsize                 & 0.109 & 0.913 & 0.010 & 0.019    & 0.489 & 0.007  & \xmark                & \xmark                      &                      &  6,358              \\
        \thickhline
        \multirow{3}{*}{\rotatebox[origin=c]{90}{\parbox[c]{1.4cm}{\centering Ours}}} & TOKI-lime \small (Ours) \normalsize   & 0.922 & 0.977 & 0.938 & 0.957    & 0.737 & 0.831  & 12,000                  & 1,126,453                       & \multirow{3}{*}{245}  &  23,892  \\
        \cline{2-10}
        \cline{12-12}
        & Naive                      & 0.861 & 0.925 & 0.925 & 0.925    & 0.105 & 0.312  & \xmark                & \xmark                      &                      & \xmark             \\
        \cline{2-10}
        \cline{12-12}
        & TOKI-lime \small (--K.I) \normalsize                 & 0.408 & 0.968 & 0.403 & 0.569    & 0.474 & 0.437  & \xmark                & \xmark                      &                      &  23,849              \\
        \hline
    \end{tabular}
}
\end{table*}

The \textit{next} experiment compares
TOKI, Naive, and TOKI (--K.I) with different explanation methods against the benchmark outlined in Table~\ref{table:ground_truths}. 
The experimental results
are presented in Table~\ref{table:experimental_result_rq2_5_6}. 
\begin{itemize}
    \item \textbf{RQ2.2}: \textit{Comparing TOKI and Naive}.
    TOKI consistently outperforms Naive across all datasets. 
    Naive occasionally shows good precision, sensitivity and F1-score but relatively low specificity and G-mean, especially on Dong's, Issues, and our trustworthiness datasets.
    This indicates that Naive is biased toward labeling predictions as trustworthy, resulting in misclassifying untrustworthy predictions as trustworthy.

    \item \textbf{RQ2.3}: \textit{Ablation study~--~the effect of keywords}.
    TOKI (--K.I) underperforms compared to TOKI with low accuracy and G-mean across all datasets.
    On Dong's and HateXplain trustworthiness datasets, TOKI (--K.I) shows high specificity and precision, but other metrics remain low.
    This suggests a bias against labeling predictions as trustworthy, leading to misclassifying trustworthy predictions as untrustworthy.
    In terms of efficiency, TOKI (–K.I) and TOKI have similar processing times for computing trustworthiness labels, indicating that most of the time is spent on explaining predictions.
    Although keyword identification is time-consuming, it is only run once to assess the trustworthiness of a batch of predictions.

    \item \textbf{RQ2.4}: \textit{The impact of explanation methods}. 
    We investigate this using Dong's, CAMS, and HateXplain datasets, as other datasets only use LIME.
    Overall, TOKI demonstrates the best effectiveness with the SOTA LIME. 
    TOKI-lime achieves higher accuracy and G-mean compared to TOKI-omis and TOKI-grad, while
    other metrics
    show no significant differences, except for the CAMS dataset.
    Explanation methods also impact TOKI’s efficiency.
    TOKI-lime is significantly slower due to LIME's sampling of 5,000 neighbors, while omission and gradient are more lightweight.
    The model complexity also affects TOKI's efficiency, as the fine-tuned models consume more time than the two simpler MLPs in Dong's dataset.
    
\end{itemize}

\begin{custombox}[Answer to RQ2]
TOKI outperforms the naive baseline based solely on model confidence, achieving the best effectiveness with LIME explanations.
Without identifying keywords, TOKI's ability to measure semantic relatedness is limited, which can misclassify trustworthy predictions as untrustworthy.
In terms of efficiency, TOKI depends on the explanation methods and the model complexity, as most of its processing time is spent on explaining predictions.
\end{custombox}

\subsubsection*{\textbf{RQ3}: Effectiveness of TOKI-guided attack method}

\begin{table}[t]
    \caption{Experimental results comparing TOKI-guided attack method and A2T~\cite{Yoo:2021:TowardsImprovingAdversarialTrainingofNLPModels} 
    }
    \label{table:epxr_attack_results}
    \resizebox{\textwidth}{!}{
    \begin{tabular}{|l|c|c|c|c|c|c|c|c|c|}
    \hline
    \multicolumn{1}{|c|}{\multirow{2}{*}{\textbf{Dataset}}} & \multicolumn{1}{c|}{\multirow{2}{*}{\makecell[c]{\textbf{Instances}}}} & \multicolumn{4}{c|}{\textbf{TOKI-guided attack method (Ours)}} &  \multicolumn{4}{c|}{\textbf{A2T}} \\
    \cline{3-10}
     & & \multicolumn{1}{l|}{\makecell[c]{\textbf{SAR}~($\uparrow$)}} & {\textbf{Avg.NP}~($\downarrow$)} & \multicolumn{1}{l|}{\makecell[c]{\makecell[c]{\textbf{Bert}~($\uparrow$)}}} & \multicolumn{1}{l|}{\makecell[c]{\makecell[c]{\textbf{USE}~($\uparrow$)}}} & \multicolumn{1}{l|}{\makecell[c]{\textbf{SAR}~($\uparrow$)}} & {\textbf{Avg.NP}~($\downarrow$)} & \multicolumn{1}{l|}{\makecell[c]{\makecell[c]{\textbf{Bert}~($\uparrow$)}}} & \multicolumn{1}{l|}{\makecell[c]{\makecell[c]{\textbf{USE}~($\uparrow$)}}} \\
     \hline 
     movies & 728 & 0.994 & \cellcolor{gray!15}{11.399} & 0.994 & 0.996  & 0.994 & 11.531 & 0.994 & 0.996  \\ 
     20news & 926 & \cellcolor{gray!15}{0.458} & \cellcolor{gray!15}{21.058} & \cellcolor{gray!15}{0.977} & 0.979 & 0.456 & 21.089 & 0.976 & 0.979 \\ 
     CAMS & 4,316 & \cellcolor{gray!15}{0.812} & \cellcolor{gray!15}{4.807} & \cellcolor{gray!15}{0.983} & \cellcolor{gray!15}{0.968} & 0.765 & 5.500 & 0.981 & 0.966 \\ 
     HateXplain & 3,846 & \cellcolor{gray!15}{0.351} & \cellcolor{gray!15}{1.887} & \cellcolor{gray!15}{0.952} & \cellcolor{gray!15}{0.933} & 0.325 & 1.915 & 0.951 & 0.931 \\ 
     \revised{Issues} & 3,090 & \cellcolor{gray!15}{0.220}  & \cellcolor{gray!15}{3.629}  & \cellcolor{gray!15}{0.967}  & \cellcolor{gray!15}{0.975}  & 0.168  & 4.255  & 0.960  & 0.972  \\ 
     amazon\_polarity & 8,000 & \cellcolor{gray!15}{0.362} & \cellcolor{gray!15}{5.976} & \cellcolor{gray!15}{0.974} & \cellcolor{gray!15}{0.969} & 0.187 & 7.316 & 0.970 & 0.963 \\ 
     ag\_news & 7,600 & \cellcolor{gray!15}{0.192} & \cellcolor{gray!15}{2.481} & \cellcolor{gray!15}{0.965} & 0.949 & 0.151 & 2.668 & 0.963 & \cellcolor{gray!15}{0.952} \\ 
     rotten\_tomatoes & 1,066 & \cellcolor{gray!15}{0.500} & \cellcolor{gray!15}{1.626} & \cellcolor{gray!15}{0.952} & \cellcolor{gray!15}{0.921} & 0.334 & 1.898 & 0.950 & 0.919 \\ 
     yahoo\_answers\_topics & 8,000 & \cellcolor{gray!15}{0.372}  & \cellcolor{gray!15}{4.107} & \cellcolor{gray!15}{0.955} & \cellcolor{gray!15}{0.954} & 0.307 & 4.726 & 0.953 & 0.952 \\ 
     imdb & 8,000 & \cellcolor{gray!15}{0.800} & \cellcolor{gray!15}{7.473} & \cellcolor{gray!15}{0.983} & \cellcolor{gray!15}{0.985} & 0.650 & 9.786 & 0.981 & 0.982 \\ 
     emotion & 2,000 & \cellcolor{gray!15}{0.643} & \cellcolor{gray!15}{1.152} & \cellcolor{gray!15}{0.961} & \cellcolor{gray!15}{0.927} & 0.517 & 1.235 & 0.957 & 0.926 \\ 
     \hline
    \end{tabular}%
    }
\end{table}
\begin{table}[t]
    \centering
    \caption{Several adversarial examples: \textcolor{blue}{{\textit{blue}}} words indicate perturbations in the text, \textcolor{green}{\cmark} and \textcolor{red}{\xmark} represent successful and failed attacks, respectively}
    \label{table:adversarial_examples}
    \resizebox{\textwidth}{!}{
    \begin{tabular}{|p{0.6\linewidth}|p{0.64\linewidth}|}
        \hline
        \textbf{Original [toxic]:} If I am a moderate muzzy and do not stand against the extremists within my religion, I am a worthless dog who deserves the title of muzzy terrorist period. 
            & \textbf{Original [medication]:} When I manage to get a small victory against depression and ``wake up'', I still have to fight against the pain that doesn't let me move without moaning. \\
        \textbf{A2T [normal~\textcolor{green}{\cmark}]:} If I am a moderate muzzy and do not stand against the extremists within my \textcolor{blue}{\textit{{cults}}}, I am a \textcolor{blue}{\textit{{pointless}}} dog who \textcolor{blue}{\textit{{merits}}} the title of muzzy terrorist \textcolor{blue}{\textit{{periods}}}.
            & \textbf{A2T [jobs and careers~\textcolor{green}{\cmark}]:} When I manage to get a small \textcolor{blue}{\textit{{win}}} against \textcolor{blue}{\textit{{downturn}}} and ``wake up'', I still have to fight against the \textcolor{blue}{\textit{{grief}}} that doesn't let me move without \textcolor{blue}{\textit{{whining}}}. \\
        \textbf{TOKI [normal~\textcolor{green}{\cmark}]:} If I am a \textcolor{blue}{\textit{{mild}}} muzzy and do not stand against the extremists within my religion, I am a worthless dog who deserves the title of muzzy terrorist period.
            & \textbf{TOKI [jobs and careers~\textcolor{green}{\cmark}]:} When I manage to get a small victory against \textcolor{blue}{\textit{{stressors}}} and ``wake up'', I still have to fight against the pain that doesn't let me move without moaning. \\
        \hline
        \textbf{Original [joy]:} He is old enough to no longer feel that I am the only acceptable answer in the dark. 
            & \textbf{Original [negative]:} Despite its visual virtuosity, ``Naqoyqatsi'' is banal in its message and the choice of material to convey it. \\
        \textbf{A2T [sadness~\textcolor{green}{\cmark}]:} He is \textcolor{blue}{\textit{{longtime}}} enough to no longer feel that I am the only \textcolor{blue}{\textit{{agreeable}}} answer in the \textcolor{blue}{\textit{{gloomy}}}. 
            & \textbf{A2T [negative~\textcolor{red}{\xmark}]:} \textcolor{blue}{\textit{{Though}}} its visual virtuosity, ``Naqoyqatsi'' is banal in its \textcolor{blue}{\textit{{messaging}}} and the \textcolor{blue}{\textit{{choices}}} of \textcolor{blue}{\textit{{materials}}} to convey it. \\
        \textbf{TOKI [love~\textcolor{green}{\cmark}]:} He is old enough to no longer feel that I am the only \textcolor{blue}{\textit{{accepted}}} answer in the dark. 
            & \textbf{TOKI [positive~\textcolor{green}{\cmark}]:} Despite its visual virtuosity, ``Naqoyqatsi'' is \textcolor{blue}{\textit{{insipid}}} in its message and the choice of material to convey it. \\
        \hline
    \end{tabular}%
    }
    \label{table:attack_example}
\end{table}

Table~\ref{table:epxr_attack_results} compares the effectiveness of TOKI-guided attack method and A2T~\cite{Yoo:2021:TowardsImprovingAdversarialTrainingofNLPModels}.
Overall, TOKI-guided attack method outperforms A2T, achieving a higher ASR and lower NP across all models.
Adversarial examples generated by TOKI also have higher Bert
and USE scores than those by A2T, except for the USE score on the ag\_news dataset, demonstrating TOKI's ability to preserve the semantic meaning of the original texts.
Table~\ref{table:adversarial_examples} displays several adversarial examples generated by both methods.
It is observed that
TOKI can effectively find synonyms to attack with fewer perturbed words, while A2T requires more perturbations and sometimes fails.
\revised{Interestingly, in Dong’s dataset, A2T performs nearly as well as the proposed method, likely because the models are simple MLPs and more easily attacked than transformer-based models in other datasets. 
However, in other datasets, the proposed method outperforms A2T by 2.6\%–17.5\% in ASR and requires 0.1–1.34 fewer NP. These results underscore the effectiveness of TOKI-guided attack method, particularly against models using SOTA architectures.}

\begin{custombox}[Answer to RQ3]
The adversarial attack method guided by TOKI outperforms the SOTA A2T.
Our method achieves a higher success rate with fewer perturbations than A2T while preserving the semantic meaning of the original texts.
\end{custombox}

\section{Discussion} 
\label{sec:expr_discussion}
\subsection{Implications}

\subsubsection*{\textbf{Prediction uncertainty and trustworthiness}.}
Experimental results show that highly confident predictions are more likely to be trustworthy.
However, high confidence does not guarantee trustworthiness, and low confidence predictions can still be trustworthy.
Relying on prediction uncertainty is ineffective for trustworthiness assessment, overlooking both trustworthy low and untrustworthy high confidence predictions.
Prior work has a complex debate over the impact of prediction uncertainty, mainly on human trust in ML models.
Several studies~\cite{Anh:2015:DeepNeuralNetworksAreEasilyFooledHighConfidencePredictionsForUnrecognizableImages, Bussone:2015:TheRoleofExplanationsonTrustandRelianceinClinicalDecisionSupportSystems} suggest that prediction uncertainty has a limited impact on human trust~\cite{Rechkemmer:2022:WhenConfidenceMeetsAccuracyExploringtheEffectsofMultiplePerformanceIndicatorsonTrustinMachineLearningModels, Ovadia:2019:CanYouTrustYourModelsUncertaintyEvaluatingPredictiveUncertaintyUnderDatasetShift}.
Other studies~\cite{Zhang:2020:EffectOfConfidenceAndExplanationOnAccuracyAndTrustCalibrationInAiAssistedDecisionMaking} show that prediction certainty can improve human trust in ML models and increase the willingness to rely on high confidence predictions.
However, we argue that whether prediction certainty increases human trust or not, it does not translate into improving the model's trustworthiness.
Model certainty can lead to overreliance on ML decisions without reflecting the true trustworthiness of the model. 
The model may make decisions based on spurious correlations, being confident in them rather than valid correlations in the training data.
This fits the distinction between trust and trustworthiness defined by \citeN{Kästner:2021:OntheRelationofTrustandExplainabilityWhytoEngineerforTrustworthiness},
where people can still trust an untrustworthy model.


\subsubsection*{\textbf{Automated trustworthiness oracles}.}
TOKI identifies keywords for each class and uses them to measure the semantic relatedness between words and the class, effectively capturing the characteristics of classes.
This makes TOKI outperform the baselines, including its variant that omits keywords to measure relatedness, which is biased for untrustworthy predictions.

Regarding adversarial attacks, examples generated by TOKI are more effective in attacking models than those created by A2T. 
By leveraging correlations between words and classes, our method generates adversarial examples likely to trigger trustworthiness vulnerabilities.
This enables TOKI-guided attack method to achieve a higher attack success rate with fewer perturbations than existing adversarial attack methods. 
This finding highlights the negative impact of trustworthiness issues and the need for trustworthiness oracles, which remain underexplored in the research community.

\subsubsection*{\textbf{\revised{Implications for software engineering}}.}
\revised{
Trustworthiness is a crucial non-functional requirement of ML systems~\cite{Riccio:2020:TestingMachineLearningBasedSystems:ASystematicMapping}.
Trustworthiness oracles enable automating trustworthiness testing, which is particularly valuable in the iterative development of ML systems. 
They support continuous integration and delivery pipelines, monitoring in real-world SE environments, such as online testing or DevOps, and handling corner-cased and real-world inputs. 
Since TOKI is efficient and lightweight, it can be directly applied in online settings and real-world environments without extensive training, requiring only a one-time keyword identification. 
This makes TOKI a practical safety net for software systems that rely on ML text classification.
TOKI's outputs also provide actionable insights for feedback-driven improvement, reducing the need for human intervention.}

\revised{Text classification is also a primary application in SE domains to automate tasks such as sentiment analysis of SE artifacts, including git commit comments, JIRA issues, and app reviews, assessing developers’ psychological states, analysing Q\&A sites like StackOverflow, software requirements classification, and project issue categorisation.
Integrating such oracles can foster greater trust in ML4SE systems and significantly advance their development and deployment.}

\subsection{Threats to Validity}
\label{sec:threats2validity}
Threats to \textbf{internal validity} can be related to the faithfulness of explanations,
which refers to how accurately the explanations reflect a model's reasoning~\cite{Mariotti:2024:TextFocus:AssessingtheFaithfulnessofFeatureAttributionMethodsExplanationsinNaturalLanguageProcessing}. 
To assess this, we evaluate TOKI using three explanation methods, including LIME, omission, and gradient. 
Another threat to internal validity is the bias in word embedding methods due to their training data.
We mitigate this by using ensemble learning with multiple word embedding methods.

To reduce threats to \textbf{external validity}, we evaluate TOKI on approximately 8,000 predictions across 11 models and datasets in various domains, including topic classification, sentiment analysis, clinical mental text classification, hate speech detection, and software issue classification.
We also employ TOKI with various explanation methods to assess its performance.

Threats to \textbf{construct validity} can arise from the trustworthiness benchmark.
We leverage existing studies, which do not directly label predictions as trustworthy or not, by measuring the explanations' plausibility.
To mitigate the threats, we collect an additional, more straightforward trustworthiness dataset.
The bias in human annotations is addressed by combining all human-annotated trustworthiness labels for each prediction using plurality voting~\cite{Leon:2017:EvaluatingTheEffectOfVotingMethodsOnEnsembleBasedClassification}.

\section{Related Work}
\label{sec:related_work}
\subsection{Machine Learning Testing}

Various testing techniques have emerged to assess different aspects of ML models.
For example, test input generation methods~\cite{Pei:2017:DeepXplore:AutomatedWhiteboxTestingofDeepLearningSystems, Guo:2018:DLFuzz:DifferentialFuzzingTestingofDeepLearningSystems, Tian:2018:DeepTest:AutomatedTestingOfDeepNeuralNetworkDrivenAutonomousCars} are used detect defects, guided by test adequacy criteria~\cite{Pei:2017:DeepXplore:AutomatedWhiteboxTestingofDeepLearningSystems, Ma:2018:DeepGauge:MultiGranularityTestingCriteriaForDeepLearningSystems}.
Several methods aim to debug and repair models~{\cite{Krishnan:2017:PALM:MachineLearningExplanationsForIterativeDebugging, Dutta:2019:Storm:ProgramReductionforTestingandDebuggingProbabilisticProgrammingSystems}.
Other methods, such as A2T~\cite{Yoo:2021:TowardsImprovingAdversarialTrainingofNLPModels}, focus on adversarial attacks~\cite{Zhang:2020:AdversarialAttacksonDeeplearningModelsinNaturalLanguageProcessingASurvey, Morris:2020:TextAttack:AFrameworkforAdversarialAttacksDataAugmentationandAdversarialTraininginNLP} to assess model robustness.
The oracle problem~{\cite{Barr:2015:TheOracleProbleminSoftwareTestingASurvey}} is also a significant area of interest. 
Two main approaches are primarily used to address the oracle problem: metamorphic testing~\cite{Ramanagopal:2018:FailingtoLearn:AutonomouslyIdentifyingPerceptionFailuresforSelfDrivingCars, Xie:2020:METTLE:AMETamorphicTestingApproachtoAssessingandValidatingUnsupervisedMachineLearningSystems} and cross-referencing~\cite{Pei:2017:DeepXplore:AutomatedWhiteboxTestingofDeepLearningSystems, Guo:2018:DLFuzz:DifferentialFuzzingTestingofDeepLearningSystems}.
Considerable efforts have been made to test various properties of ML models, such as accuracy, relevance, efficiency, robustness, fairness, and interpretability~{\cite{Zhang:2022:MachineLearningTestingSurveyLandscapesandHorizons}}.
However, testing the trustworthiness of ML models remains a significant challenge.
Several metrics have been proposed to quantify the concept of trustworthiness~\cite{Kaur:2021:TrustworthyExplainabilityAcceptanceANewMetrictoMeasuretheTrustworthinessofInterpretableAIMedicalDiagnosticSystems, Cheng:2020:ThereIsHopeAfterAllQuantifyingOpinionandTrustworthinessinNeuralNetworks}. However, different from other properties, much of the work on trustworthiness testing relies on human interpretation within their systems.


\subsection{\revised{Explanation Method}}

\revised{Local explanations, in contrast to global counterparts, focus on how a specific input leads to a prediction. 
This makes them particularly suitable for understanding ML predictions. 
In text classification, various methods have been developed to generate local explanations.}

\revised{\textit{Feature attribution-based} explanations assess the relevance of individual input features, such as words, to a model’s prediction. Some~\cite{Li:2017:UnderstandingNeuralNetworksthroughRepresentationErasure, Dong:2018:ComparingAutomaticandHumanEvaluationofLocalExplanationsforTextClassification} achieve this by removing, masking, or modifying input features, and observing the impact on the prediction. 
Others calculate the gradient of the output with respect to the input~\cite{Mohebbi:2021:ExploringtheRoleofBERTTokenRepresentationstoExplainSentenceProbingResults}.
Methods like LIME~\cite{Ribeiro:2016:WhyShouldITrustYouExplainingthePredictionsofAnyClassifier} approximate the model's behaviour with simpler, interpretable models.
}

\revised{\textit{Attention-based} explanations leverage the attention mechanism, which often serves as a means to attend to the most relevant part of inputs~\cite{Wiegreffe:2019:AttentionIsNotNotExplanation}. Intuitively, they can capture meaningful correlations between intermediate states of the inputs, potentially explaining the model’s predictions~\cite{Zhao:2024:ExplainabilityforLargeLanguageModelsASurvey}. Hence, many approaches~\cite{Yeh:2024:AttentionViz:AGlobalViewofTransformerAttention, Barkan:2021:GradSAM:ExplainingTransformersviaGradientSelfAttentionMaps} aim to explain the predictions solely based on the attention weights or by analysing the knowledge encoded in the attention.}

\revised{
\textit{Counterfactual} explanations reveal model behavior by demonstrating how slight input changes impact outputs, highlighting key features influencing predictions~\cite{Zhao:2024:ExplainabilityforLargeLanguageModelsASurvey}. Originally applied to ML tasks with explicit features and tabular datasets~\cite{Guidotti:2024:CounterfactualExplanationsAndHowToFindThem:LiteratureReviewAndBenchmarking}, these explanations have since been extended to other tasks, including text classification~\cite{Treviso:2023:CREST:AJointFrameworkforRationalizationandCounterfactualTextGeneration}.}


\revised{While other local explanation methods are subject to extensive debate~\cite{Jain:2019:AttentionisnotExplanation}, feature attribution explanations, such as the SOTA LIME~\cite{Ribeiro:2016:WhyShouldITrustYouExplainingthePredictionsofAnyClassifier}, have proven to be effective, faithful~\cite{Mariotti:2024:TextFocus:AssessingtheFaithfulnessofFeatureAttributionMethodsExplanationsinNaturalLanguageProcessing}, and widely used by AI practitioners.
Therefore, this paper focuses on feature attribution-based explanations.
Exploring other local explanations for trustworthiness testing is a promising direction, which we leave for future work.
}

\subsection{Making Use of Explanations}
Explanations have served as a valuable tool for testing and improving ML models~\cite{Zhao:2024:ExplainabilityforLargeLanguageModelsASurvey}.
Various studies have used explanations to understand and debug models.
\citeN{Ribeiro:2016:WhyShouldITrustYouExplainingthePredictionsofAnyClassifier} demonstrated the usefulness of local explanations in various tasks, such as model comparison and trust assessment.
\citeN{Lapuschkin:2019:UnmaskingCleverHansPredictorsAndAssessingWhatMachinesReallyLearn} introduced a semi-automated approach that characterises and validates classification strategies. 
\citeN{Thomas:2019:AnalyzingNeuroimagingDataThroughRecurrentDeepLearningModels} used explanations to uncover input-prediction patterns.
Several studies~\cite{Yoo:2021:TowardsImprovingAdversarialTrainingofNLPModels, Zhang:2020:AdversarialAttacksonDeeplearningModelsinNaturalLanguageProcessingASurvey} have applied explanations in adversarial attacks, primarily to determine word substitution order based on word importance.
To the best of our knowledge, this paper introduces the first approach that leverages explanations to make perturbations, specifically, by weakening valid correlations and strengthening spurious ones.
Explanations have also been integrated into the learning process to improve model performance and reliability.
\citeN{Andrew:2017:RightfortheRightReasonsTrainingDifferentiableModelsbyConstrainingtheirExplanations} proposed an approach to discover multiple models for the same task with different classification strategies, allowing domain experts to choose the best one. 
\citeN{Chen:2022:AdversarialTrainingforImprovingModelRobustnessLookatBothPredictionandInterpretation} used explanations to improve the adversarial robustness of language models.
Other studies~\cite{Schramowski:2020:MakingDeepNeuralNetworksRightForTheRightScientificReasonsByInteractingWithTheirExplanations, Ghai:2020:ExplainableActiveLearningXALTowardAIExplanationsasInterfacesforMachineTeachers} 
leveraged explanations to make models right for the right scientific reasons.
Similarly, 
\citeN{Du::2021:TowardsInterpretingandMitigatingShortcutLearningBehaviorofNLUmodels} developed a framework to mitigate shortcuts, focusing on stop words, punctuation, and numbers.
Recent studies~\cite{Linhardt:2024:PreemptivelyPruningCleverHansStrategiesinDeepNeuralNetworks} have also applied explanation regularisation to mitigate the impact of spurious correlations.


\section{Conclusion}
\label{sec:conclusion}
We investigate the trustworthiness oracle problem of text classifiers. 
Statistical evaluation reveals that while highly confident predictions are more likely to be trustworthy, some still lack trustworthiness due to reliance on spurious correlations. 
We propose TOKI, an automated trustworthiness oracle generation method. 
Experiments compare the effectiveness and efficiency of TOKI, TOKI's ablation variant and the naive approach against the trustworthiness benchmark on ten models and datasets. 
Results show that TOKI outperforms other approaches.
We also introduce a TOKI-guided adversarial attack method, which proves to be more effective than the SOTA A2T. 
In addition, several directions remain open for future work. 
We plan to explore alternative explanations, such as attention-based and counterfactual explanations, as well as other word embedding methods like sentence-transformers~\cite{Reimers:2019:SentenceBERT:SentenceEmbeddingsusingSiameseBERTNetworks}.
Further experiments on SE-specific datasets are also necessary to gain more insights into the trustworthiness problem in SE.
Moreover, adapting TOKI to other data types, such as images and speech, 
presents a promising avenue in the future.

\subsubsection*{\textbf{Data Availability}.}
Our data and replicate package are available at~\cite{Ours:2024:Zenodo}.

\bibliographystyle{ACM-Reference-Format}
\bibliography{sample-base}


\begin{thebibliography}{82}


\ifx \showCODEN    \undefined \def \showCODEN     #1{\unskip}     \fi
\ifx \showISBNx    \undefined \def \showISBNx     #1{\unskip}     \fi
\ifx \showISBNxiii \undefined \def \showISBNxiii  #1{\unskip}     \fi
\ifx \showISSN     \undefined \def \showISSN      #1{\unskip}     \fi
\ifx \showLCCN     \undefined \def \showLCCN      #1{\unskip}     \fi
\ifx \shownote     \undefined \def \shownote      #1{#1}          \fi
\ifx \showarticletitle \undefined \def \showarticletitle #1{#1}   \fi
\ifx \showURL      \undefined \def \showURL       {\relax}        \fi
\providecommand\bibfield[2]{#2}
\providecommand\bibinfo[2]{#2}
\providecommand\natexlab[1]{#1}
\providecommand\showeprint[2][]{arXiv:#2}

\bibitem[Adadi and Berrada(2018)]%
        {Adadi:2018:PeekingInsidetheBlackBoxASurveyonExplainableArtificialIntelligenceXAI}
\bibfield{author}{\bibinfo{person}{Amina Adadi} {and} \bibinfo{person}{Mohammed Berrada}.} \bibinfo{year}{2018}\natexlab{}.
\newblock \showarticletitle{Peeking Inside the Black-Box: A Survey on Explainable Artificial Intelligence (XAI)}.
\newblock \bibinfo{journal}{\emph{IEEE Access}}  \bibinfo{volume}{6} (\bibinfo{year}{2018}), \bibinfo{pages}{52138--52160}.
\newblock
\href{https://doi.org/10.1109/ACCESS.2018.2870052}{doi:\nolinkurl{10.1109/ACCESS.2018.2870052}}


\bibitem[Barkan et~al\mbox{.}(2021)]%
        {Barkan:2021:GradSAM:ExplainingTransformersviaGradientSelfAttentionMaps}
\bibfield{author}{\bibinfo{person}{Oren Barkan}, \bibinfo{person}{Edan Hauon}, \bibinfo{person}{Avi Caciularu}, \bibinfo{person}{Ori Katz}, \bibinfo{person}{Itzik Malkiel}, \bibinfo{person}{Omri Armstrong}, {and} \bibinfo{person}{Noam Koenigstein}.} \bibinfo{year}{2021}\natexlab{}.
\newblock \showarticletitle{Grad-SAM: Explaining Transformers via Gradient Self-Attention Maps}. In \bibinfo{booktitle}{\emph{ACM International Conference on Information \& Knowledge Management}} (Queensland, Australia). \bibinfo{pages}{2882–2887}.
\newblock
\showISBNx{9781450384469}
\href{https://doi.org/10.1145/3459637.3482126}{doi:\nolinkurl{10.1145/3459637.3482126}}


\bibitem[Barr et~al\mbox{.}(2015)]%
        {Barr:2015:TheOracleProbleminSoftwareTestingASurvey}
\bibfield{author}{\bibinfo{person}{Earl~T. Barr}, \bibinfo{person}{Mark Harman}, \bibinfo{person}{Phil McMinn}, \bibinfo{person}{Muzammil Shahbaz}, {and} \bibinfo{person}{Shin Yoo}.} \bibinfo{year}{2015}\natexlab{}.
\newblock \showarticletitle{The Oracle Problem in Software Testing: A Survey}.
\newblock \bibinfo{journal}{\emph{IEEE Trans. on Soft. Eng.}} \bibinfo{volume}{41}, \bibinfo{number}{5} (\bibinfo{year}{2015}), \bibinfo{pages}{507--525}.
\newblock
\href{https://doi.org/10.1109/TSE.2014.2372785}{doi:\nolinkurl{10.1109/TSE.2014.2372785}}


\bibitem[Bengio et~al\mbox{.}(2000)]%
        {Bengio:2000:ANeuralProbabilisticLanguageModel}
\bibfield{author}{\bibinfo{person}{Yoshua Bengio}, \bibinfo{person}{R\'{e}jean Ducharme}, {and} \bibinfo{person}{Pascal Vincent}.} \bibinfo{year}{2000}\natexlab{}.
\newblock \showarticletitle{A Neural Probabilistic Language Model}. In \bibinfo{booktitle}{\emph{Advances in Neural Information Processing Systems}}, Vol.~\bibinfo{volume}{13}. \bibinfo{publisher}{MIT Press}.
\newblock


\bibitem[Bojanowski et~al\mbox{.}(2017)]%
        {Bojanowski:2017:FastText:EnrichingWordVectorswithSubwordInformation}
\bibfield{author}{\bibinfo{person}{Piotr Bojanowski}, \bibinfo{person}{Edouard Grave}, \bibinfo{person}{Armand Joulin}, {and} \bibinfo{person}{Tomas Mikolov}.} \bibinfo{year}{2017}\natexlab{}.
\newblock \showarticletitle{{Enriching Word Vectors with Subword Information}}.
\newblock \bibinfo{journal}{\emph{Trans. of the Assoc. for Comp. Ling.}}  \bibinfo{volume}{5} (\bibinfo{date}{06} \bibinfo{year}{2017}), \bibinfo{pages}{135--146}.
\newblock
\showISSN{2307-387X}
\href{https://doi.org/10.1162/tacl_a_00051}{doi:\nolinkurl{10.1162/tacl_a_00051}}


\bibitem[Budanitsky and Hirst(2006)]%
        {Budanitsky:2006:EvaluatingWordNetbasedMeasuresofLexicalSemanticRelatedness}
\bibfield{author}{\bibinfo{person}{Alexander Budanitsky} {and} \bibinfo{person}{Graeme Hirst}.} \bibinfo{year}{2006}\natexlab{}.
\newblock \showarticletitle{{Evaluating WordNet-based Measures of Lexical Semantic Relatedness}}.
\newblock \bibinfo{journal}{\emph{Computational Linguistics}} \bibinfo{volume}{32}, \bibinfo{number}{1} (\bibinfo{date}{03} \bibinfo{year}{2006}), \bibinfo{pages}{13--47}.
\newblock
\showISSN{0891-2017}
\href{https://doi.org/10.1162/coli.2006.32.1.13}{doi:\nolinkurl{10.1162/coli.2006.32.1.13}}


\bibitem[Bussone et~al\mbox{.}(2015)]%
        {Bussone:2015:TheRoleofExplanationsonTrustandRelianceinClinicalDecisionSupportSystems}
\bibfield{author}{\bibinfo{person}{Adrian Bussone}, \bibinfo{person}{Simone Stumpf}, {and} \bibinfo{person}{Dympna O'Sullivan}.} \bibinfo{year}{2015}\natexlab{}.
\newblock \showarticletitle{The Role of Explanations on Trust and Reliance in Clinical Decision Support Systems}. In \bibinfo{booktitle}{\emph{Healthcare Informatics}}. \bibinfo{pages}{160--169}.
\newblock
\href{https://doi.org/10.1109/ICHI.2015.26}{doi:\nolinkurl{10.1109/ICHI.2015.26}}


\bibitem[Canbek et~al\mbox{.}(2022)]%
        {Canbek:2022:PToPIA:ComprehensiveReviewAnalysisandKnowledgeRepresentationofBinaryClassificationPerformanceMeasuresMetrics}
\bibfield{author}{\bibinfo{person}{G{\"u}rol Canbek}, \bibinfo{person}{Tugba Taskaya~Temizel}, {and} \bibinfo{person}{Seref Sagiroglu}.} \bibinfo{year}{2022}\natexlab{}.
\newblock \showarticletitle{PToPI: A Comprehensive Review, Analysis, and Knowledge Representation of Binary Classification Performance Measures/Metrics}.
\newblock \bibinfo{journal}{\emph{SN Computer Science}} \bibinfo{volume}{4}, \bibinfo{number}{1} (\bibinfo{date}{16 Oct} \bibinfo{year}{2022}), \bibinfo{pages}{13}.
\newblock
\showISSN{2661-8907}
\href{https://doi.org/10.1007/s42979-022-01409-1}{doi:\nolinkurl{10.1007/s42979-022-01409-1}}


\bibitem[Caruana et~al\mbox{.}(2015)]%
        {Caruana:2015:IntelligibleModelsforHealthCarePredictingPneumoniaRiskandHospital30dayReadmission}
\bibfield{author}{\bibinfo{person}{Rich Caruana}, \bibinfo{person}{Yin Lou}, \bibinfo{person}{Johannes Gehrke}, \bibinfo{person}{Paul Koch}, \bibinfo{person}{Marc Sturm}, {and} \bibinfo{person}{Noemie Elhadad}.} \bibinfo{year}{2015}\natexlab{}.
\newblock \showarticletitle{Intelligible Models for HealthCare: Predicting Pneumonia Risk and Hospital 30-day Readmission}. In \bibinfo{booktitle}{\emph{The 21th ACM SIGKDD International Conference on Knowledge Discovery and Data Mining}} \emph{(\bibinfo{series}{KDD '15})}. \bibinfo{publisher}{Association for Computing Machinery}, \bibinfo{address}{New York, USA}, \bibinfo{pages}{1721–1730}.
\newblock
\showISBNx{9781450336642}
\href{https://doi.org/10.1145/2783258.2788613}{doi:\nolinkurl{10.1145/2783258.2788613}}


\bibitem[Cer et~al\mbox{.}(2018)]%
        {Cer:2018:UniversalSentenceEncoderforEnglish}
\bibfield{author}{\bibinfo{person}{Daniel Cer}, \bibinfo{person}{Yinfei Yang}, \bibinfo{person}{Sheng-yi Kong}, \bibinfo{person}{Nan Hua}, \bibinfo{person}{Nicole Limtiaco}, \bibinfo{person}{Rhomni St.~John}, \bibinfo{person}{Noah Constant}, \bibinfo{person}{Mario Guajardo-Cespedes}, \bibinfo{person}{Steve Yuan}, \bibinfo{person}{Chris Tar}, \bibinfo{person}{Brian Strope}, {and} \bibinfo{person}{Ray Kurzweil}.} \bibinfo{year}{2018}\natexlab{}.
\newblock \showarticletitle{Universal Sentence Encoder for {E}nglish}. In \bibinfo{booktitle}{\emph{2018 Conference on Empirical Methods in Natural Language Processing: System Demonstrations}}. \bibinfo{publisher}{Association for Computational Linguistics}, \bibinfo{address}{Brussels, Belgium}, \bibinfo{pages}{169--174}.
\newblock
\href{https://doi.org/10.18653/v1/D18-2029}{doi:\nolinkurl{10.18653/v1/D18-2029}}


\bibitem[Chen and Ji(2022)]%
        {Chen:2022:AdversarialTrainingforImprovingModelRobustnessLookatBothPredictionandInterpretation}
\bibfield{author}{\bibinfo{person}{Hanjie Chen} {and} \bibinfo{person}{Yangfeng Ji}.} \bibinfo{year}{2022}\natexlab{}.
\newblock \showarticletitle{Adversarial Training for Improving Model Robustness? Look at Both Prediction and Interpretation}.
\newblock \bibinfo{journal}{\emph{The AAAI Conference on Artificial Intelligence}} \bibinfo{volume}{36}, \bibinfo{number}{10} (\bibinfo{date}{Jun.} \bibinfo{year}{2022}), \bibinfo{pages}{10463--10472}.
\newblock
\href{https://doi.org/10.1609/aaai.v36i10.21289}{doi:\nolinkurl{10.1609/aaai.v36i10.21289}}


\bibitem[Cheng et~al\mbox{.}(2020)]%
        {Cheng:2020:ThereIsHopeAfterAllQuantifyingOpinionandTrustworthinessinNeuralNetworks}
\bibfield{author}{\bibinfo{person}{Mingxi Cheng}, \bibinfo{person}{Shahin Nazarian}, {and} \bibinfo{person}{Paul Bogdan}.} \bibinfo{year}{2020}\natexlab{}.
\newblock \showarticletitle{There Is Hope After All: Quantifying Opinion and Trustworthiness in Neural Networks}.
\newblock \bibinfo{journal}{\emph{Frontiers in Artificial Intelligence}}  \bibinfo{volume}{3} (\bibinfo{year}{2020}).
\newblock
\showISSN{2624-8212}
\href{https://doi.org/10.3389/frai.2020.00054}{doi:\nolinkurl{10.3389/frai.2020.00054}}


\bibitem[Cho et~al\mbox{.}(2024)]%
        {Cho:2024:TOWER}
\bibfield{author}{\bibinfo{person}{Steven Cho}, \bibinfo{person}{Seaton Cousins-Baxter}, \bibinfo{person}{Stefano Ruberto}, {and} \bibinfo{person}{Valerio Terragni}.} \bibinfo{year}{2024}\natexlab{}.
\newblock \showarticletitle{Automated Trustworthiness Testing for Machine Learning Classifiers}.
\newblock  (\bibinfo{year}{2024}).
\newblock
\showeprint[arxiv]{2406.05251}~[cs.LG]
\urldef\tempurl%
\url{https://arxiv.org/abs/2406.05251}
\showURL{%
\tempurl}


\bibitem[Devlin et~al\mbox{.}(2019)]%
        {Devlin::2019:BERTPretrainingofDeepBidirectionalTransformersforLanguageUnderstanding}
\bibfield{author}{\bibinfo{person}{Jacob Devlin}, \bibinfo{person}{Ming-Wei Chang}, \bibinfo{person}{Kenton Lee}, {and} \bibinfo{person}{Kristina Toutanova}.} \bibinfo{year}{2019}\natexlab{}.
\newblock \showarticletitle{{BERT}: Pre-training of Deep Bidirectional Transformers for Language Understanding}. In \bibinfo{booktitle}{\emph{Conference of the North {A}merican Chapter of the Association for Computational Linguistics: Human Language Technologies}}, Vol.~\bibinfo{volume}{1}. \bibinfo{address}{Minnesota}, \bibinfo{pages}{4171--4186}.
\newblock
\href{https://doi.org/10.18653/v1/N19-1423}{doi:\nolinkurl{10.18653/v1/N19-1423}}


\bibitem[Dictionary(2002)]%
        {Dictionary:2002:MerriamWebster}
\bibfield{author}{\bibinfo{person}{Merriam-Webster Dictionary}.} \bibinfo{year}{2002}\natexlab{}.
\newblock \showarticletitle{Merriam-webster}.
\newblock \bibinfo{journal}{\emph{On-line at http://www. mw. com/home. htm}} \bibinfo{volume}{8}, \bibinfo{number}{2} (\bibinfo{year}{2002}), \bibinfo{pages}{23}.
\newblock


\bibitem[Dong(2018)]%
        {Dong:2018:ComparingAutomaticandHumanEvaluationofLocalExplanationsforTextClassification}
\bibfield{author}{\bibinfo{person}{Nguyen Dong}.} \bibinfo{year}{2018}\natexlab{}.
\newblock \showarticletitle{Comparing Automatic and Human Evaluation of Local Explanations for Text Classification}. In \bibinfo{booktitle}{\emph{The 16th Annual Conference of the North American Chapter of the Association for Computational Linguistics: Human Language Technologies}}. \bibinfo{address}{Hyatt Regency, New Orleans}, \bibinfo{pages}{1069–1078}.
\newblock
\href{https://doi.org/10.18653/v1/N18-1097}{doi:\nolinkurl{10.18653/v1/N18-1097}}


\bibitem[Du et~al\mbox{.}(2023)]%
        {Du:2023:ShortcutLearningofLargeLanguageModelsinNaturalLanguageUnderstanding}
\bibfield{author}{\bibinfo{person}{Mengnan Du}, \bibinfo{person}{Fengxiang He}, \bibinfo{person}{Na Zou}, \bibinfo{person}{Dacheng Tao}, {and} \bibinfo{person}{Xia Hu}.} \bibinfo{year}{2023}\natexlab{}.
\newblock \showarticletitle{Shortcut Learning of Large Language Models in Natural Language Understanding}.
\newblock \bibinfo{journal}{\emph{Commun. ACM}} \bibinfo{volume}{67}, \bibinfo{number}{1} (\bibinfo{date}{dec} \bibinfo{year}{2023}), \bibinfo{pages}{110–120}.
\newblock
\showISSN{0001-0782}
\href{https://doi.org/10.1145/3596490}{doi:\nolinkurl{10.1145/3596490}}


\bibitem[Du et~al\mbox{.}(2021)]%
        {Du::2021:TowardsInterpretingandMitigatingShortcutLearningBehaviorofNLUmodels}
\bibfield{author}{\bibinfo{person}{Mengnan Du}, \bibinfo{person}{Varun Manjunatha}, \bibinfo{person}{Rajiv Jain}, \bibinfo{person}{Ruchi Deshpande}, \bibinfo{person}{Franck Dernoncourt}, \bibinfo{person}{Jiuxiang Gu}, \bibinfo{person}{Tong Sun}, {and} \bibinfo{person}{Xia Hu}.} \bibinfo{year}{2021}\natexlab{}.
\newblock \showarticletitle{Towards Interpreting and Mitigating Shortcut Learning Behavior of {NLU} models}. In \bibinfo{booktitle}{\emph{The North American Chapter of the Association for Computational Linguistics: Human Language Technologies}}. \bibinfo{pages}{915--929}.
\newblock
\href{https://doi.org/10.18653/v1/2021.naacl-main.71}{doi:\nolinkurl{10.18653/v1/2021.naacl-main.71}}


\bibitem[Dutta et~al\mbox{.}(2019)]%
        {Dutta:2019:Storm:ProgramReductionforTestingandDebuggingProbabilisticProgrammingSystems}
\bibfield{author}{\bibinfo{person}{Saikat Dutta}, \bibinfo{person}{Wenxian Zhang}, \bibinfo{person}{Zixin Huang}, {and} \bibinfo{person}{Sasa Misailovic}.} \bibinfo{year}{2019}\natexlab{}.
\newblock \showarticletitle{Storm: program reduction for testing and debugging probabilistic programming systems}. In \bibinfo{booktitle}{\emph{ACM Joint Meeting on European Software Engineering Conference and Symposium on the Foundations of Software Engineering}} (Estonia). \bibinfo{publisher}{Association for Computing Machinery}, \bibinfo{address}{New York, USA}, \bibinfo{pages}{729–739}.
\newblock
\showISBNx{9781450355728}
\href{https://doi.org/10.1145/3338906.3338972}{doi:\nolinkurl{10.1145/3338906.3338972}}


\bibitem[Garg et~al\mbox{.}(2022)]%
        {Garg:2022:CAMS:AnAnnotatedCorpusforCausalAnalysisofMentalHealthIssuesinSocialMediaPosts}
\bibfield{author}{\bibinfo{person}{Muskan Garg}, \bibinfo{person}{Chandni Saxena}, \bibinfo{person}{Sriparna Saha}, \bibinfo{person}{Veena Krishnan}, \bibinfo{person}{Ruchi Joshi}, {and} \bibinfo{person}{Vijay Mago}.} \bibinfo{year}{2022}\natexlab{}.
\newblock \showarticletitle{{CAMS}: An Annotated Corpus for Causal Analysis of Mental Health Issues in Social Media Posts}. In \bibinfo{booktitle}{\emph{Language Resources and Evaluation}}. \bibinfo{publisher}{European Language Resources Association}, \bibinfo{address}{France}, \bibinfo{pages}{387--396}.
\newblock
\urldef\tempurl%
\url{https://aclanthology.org/2022.lrec-1.686}
\showURL{%
\tempurl}


\bibitem[Geirhos et~al\mbox{.}(2020)]%
        {Geirhos:2020:ShortcutLearninginDeepNeuralNetworks}
\bibfield{author}{\bibinfo{person}{Robert Geirhos}, \bibinfo{person}{J{\"o}rn-Henrik Jacobsen}, \bibinfo{person}{Claudio Michaelis}, \bibinfo{person}{Richard Zemel}, \bibinfo{person}{Wieland Brendel}, \bibinfo{person}{Matthias Bethge}, {and} \bibinfo{person}{Felix~A. Wichmann}.} \bibinfo{year}{2020}\natexlab{}.
\newblock \showarticletitle{Shortcut learning in deep neural networks}.
\newblock \bibinfo{journal}{\emph{Nature Machine Intelligence}} \bibinfo{volume}{2}, \bibinfo{number}{11} (\bibinfo{date}{01 Nov} \bibinfo{year}{2020}), \bibinfo{pages}{665--673}.
\newblock
\showISSN{2522-5839}
\href{https://doi.org/10.1038/s42256-020-00257-z}{doi:\nolinkurl{10.1038/s42256-020-00257-z}}


\bibitem[Ghai et~al\mbox{.}(2021)]%
        {Ghai:2020:ExplainableActiveLearningXALTowardAIExplanationsasInterfacesforMachineTeachers}
\bibfield{author}{\bibinfo{person}{Bhavya Ghai}, \bibinfo{person}{Q.~Vera Liao}, \bibinfo{person}{Yunfeng Zhang}, \bibinfo{person}{Rachel Bellamy}, {and} \bibinfo{person}{Klaus Mueller}.} \bibinfo{year}{2021}\natexlab{}.
\newblock \showarticletitle{Explainable Active Learning (XAL): Toward AI Explanations as Interfaces for Machine Teachers}.
\newblock \bibinfo{journal}{\emph{Proc. ACM Hum.-Comput. Interact.}} \bibinfo{volume}{4}, \bibinfo{number}{CSCW3}, Article \bibinfo{articleno}{235} (\bibinfo{date}{jan} \bibinfo{year}{2021}), \bibinfo{numpages}{28}~pages.
\newblock
\href{https://doi.org/10.1145/3432934}{doi:\nolinkurl{10.1145/3432934}}


\bibitem[Guidotti(2024)]%
        {Guidotti:2024:CounterfactualExplanationsAndHowToFindThem:LiteratureReviewAndBenchmarking}
\bibfield{author}{\bibinfo{person}{Riccardo Guidotti}.} \bibinfo{year}{2024}\natexlab{}.
\newblock \showarticletitle{Counterfactual explanations and how to find them: literature review and benchmarking}.
\newblock \bibinfo{journal}{\emph{Data Mining and Knowledge Discovery}} \bibinfo{volume}{38}, \bibinfo{number}{5} (\bibinfo{date}{01 Sep} \bibinfo{year}{2024}), \bibinfo{pages}{2770--2824}.
\newblock
\showISSN{1573-756X}
\href{https://doi.org/10.1007/s10618-022-00831-6}{doi:\nolinkurl{10.1007/s10618-022-00831-6}}


\bibitem[Guo et~al\mbox{.}(2018)]%
        {Guo:2018:DLFuzz:DifferentialFuzzingTestingofDeepLearningSystems}
\bibfield{author}{\bibinfo{person}{Jianmin Guo}, \bibinfo{person}{Yu Jiang}, \bibinfo{person}{Yue Zhao}, \bibinfo{person}{Quan Chen}, {and} \bibinfo{person}{Jiaguang Sun}.} \bibinfo{year}{2018}\natexlab{}.
\newblock \showarticletitle{DLFuzz: differential fuzzing testing of deep learning systems}. In \bibinfo{booktitle}{\emph{ACM Joint Meeting on European Software Engineering Conference and Symposium on the Foundations of Software Engineering}}. \bibinfo{address}{New York, USA}, \bibinfo{pages}{739–743}.
\newblock
\showISBNx{9781450355735}
\href{https://doi.org/10.1145/3236024.3264835}{doi:\nolinkurl{10.1145/3236024.3264835}}


\bibitem[Guzman and Bruegge(2013)]%
        {Guzman2013:TowardsEmotionalAwarenessInSoftwareDevelopmentTeams}
\bibfield{author}{\bibinfo{person}{Emitza Guzman} {and} \bibinfo{person}{Bernd Bruegge}.} \bibinfo{year}{2013}\natexlab{}.
\newblock \showarticletitle{Towards emotional awareness in software development teams}. In \bibinfo{booktitle}{\emph{Joint Meeting on Foundations of Software Engineering}} (Saint Petersburg, Russia). \bibinfo{publisher}{Association for Computing Machinery}, \bibinfo{address}{New York, NY, USA}, \bibinfo{pages}{671–674}.
\newblock
\showISBNx{9781450322379}
\href{https://doi.org/10.1145/2491411.2494578}{doi:\nolinkurl{10.1145/2491411.2494578}}


\bibitem[Harrigian et~al\mbox{.}(2020)]%
        {Harrigian:2020:DoModelsofMentalHealthBasedonSocialMediaDataGeneralize}
\bibfield{author}{\bibinfo{person}{Keith Harrigian}, \bibinfo{person}{Carlos Aguirre}, {and} \bibinfo{person}{Mark Dredze}.} \bibinfo{year}{2020}\natexlab{}.
\newblock \showarticletitle{Do Models of Mental Health Based on Social Media Data Generalize?}. In \bibinfo{booktitle}{\emph{Findings of the Association for Computational Linguistics: EMNLP}}. \bibinfo{pages}{3774--3788}.
\newblock
\href{https://doi.org/10.18653/v1/2020.findings-emnlp.337}{doi:\nolinkurl{10.18653/v1/2020.findings-emnlp.337}}


\bibitem[Jain and Wallace(2019)]%
        {Jain:2019:AttentionisnotExplanation}
\bibfield{author}{\bibinfo{person}{Sarthak Jain} {and} \bibinfo{person}{Byron~C. Wallace}.} \bibinfo{year}{2019}\natexlab{}.
\newblock \showarticletitle{{A}ttention is not {E}xplanation}. In \bibinfo{booktitle}{\emph{2019 Conference of the North {A}merican Chapter of the Association for Computational Linguistics: Human Language Technologies}}, Vol.~\bibinfo{volume}{1}. \bibinfo{address}{Minnesota}, \bibinfo{pages}{3543--3556}.
\newblock
\href{https://doi.org/10.18653/v1/N19-1357}{doi:\nolinkurl{10.18653/v1/N19-1357}}


\bibitem[Kaur et~al\mbox{.}(2021)]%
        {Kaur:2021:TrustworthyExplainabilityAcceptanceANewMetrictoMeasuretheTrustworthinessofInterpretableAIMedicalDiagnosticSystems}
\bibfield{author}{\bibinfo{person}{Davinder Kaur}, \bibinfo{person}{Suleyman Uslu}, \bibinfo{person}{Arjan Durresi}, \bibinfo{person}{Sunil Badve}, {and} \bibinfo{person}{Murat Dundar}.} \bibinfo{year}{2021}\natexlab{}.
\newblock \showarticletitle{Trustworthy Explainability Acceptance: A New Metric to Measure the Trustworthiness of Interpretable AI Medical Diagnostic Systems}. In \bibinfo{booktitle}{\emph{Complex, Intelligent and Software Intensive Systems}}. \bibinfo{publisher}{Springer International Publishing}, \bibinfo{address}{Cham}, \bibinfo{pages}{35--46}.
\newblock
\showISBNx{978-3-030-79725-6}


\bibitem[Krishnan and Wu(2017)]%
        {Krishnan:2017:PALM:MachineLearningExplanationsForIterativeDebugging}
\bibfield{author}{\bibinfo{person}{Sanjay Krishnan} {and} \bibinfo{person}{Eugene Wu}.} \bibinfo{year}{2017}\natexlab{}.
\newblock \showarticletitle{PALM: Machine Learning Explanations For Iterative Debugging}. In \bibinfo{booktitle}{\emph{The 2nd Workshop on Human-In-the-Loop Data Analytics}} (Chicago, IL, USA) \emph{(\bibinfo{series}{HILDA '17})}. \bibinfo{publisher}{Association for Computing Machinery}, \bibinfo{address}{New York, USA}, Article \bibinfo{articleno}{4}, \bibinfo{numpages}{6}~pages.
\newblock
\showISBNx{9781450350297}
\href{https://doi.org/10.1145/3077257.3077271}{doi:\nolinkurl{10.1145/3077257.3077271}}


\bibitem[Kästner et~al\mbox{.}(2021)]%
        {Kästner:2021:OntheRelationofTrustandExplainabilityWhytoEngineerforTrustworthiness}
\bibfield{author}{\bibinfo{person}{Lena Kästner}, \bibinfo{person}{Markus Langer}, \bibinfo{person}{Veronika Lazar}, \bibinfo{person}{Astrid Schomäcker}, \bibinfo{person}{Timo Speith}, {and} \bibinfo{person}{Sarah Sterz}.} \bibinfo{year}{2021}\natexlab{}.
\newblock \showarticletitle{On the Relation of Trust and Explainability: Why to Engineer for Trustworthiness}. In \bibinfo{booktitle}{\emph{IEEE 29th International Requirements Engineering Conference Workshops}}. \bibinfo{pages}{169--175}.
\newblock
\href{https://doi.org/10.1109/REW53955.2021.00031}{doi:\nolinkurl{10.1109/REW53955.2021.00031}}


\bibitem[Lam et~al\mbox{.}(2024)]%
        {Ours:2024:Zenodo}
\bibfield{author}{\bibinfo{person}{Nguyen~Tung Lam}, \bibinfo{person}{Cho Steven}, \bibinfo{person}{Du Xiaoning}, \bibinfo{person}{Neelofar}, \bibinfo{person}{Terragni Valerio}, \bibinfo{person}{Ruberto Stefano}, {and} \bibinfo{person}{Aleti Aldeida}.} \bibinfo{year}{2024}\natexlab{}.
\newblock \bibinfo{title}{TOKI's Replicate Package}.
\newblock
\href{https://doi.org/10.5281/zenodo.13751579}{doi:\nolinkurl{10.5281/zenodo.13751579}}


\bibitem[Lapuschkin et~al\mbox{.}(2019)]%
        {Lapuschkin:2019:UnmaskingCleverHansPredictorsAndAssessingWhatMachinesReallyLearn}
\bibfield{author}{\bibinfo{person}{Sebastian Lapuschkin}, \bibinfo{person}{Stephan W{\"a}ldchen}, \bibinfo{person}{Alexander Binder}, \bibinfo{person}{Gr{\'e}goire Montavon}, \bibinfo{person}{Wojciech Samek}, {and} \bibinfo{person}{Klaus-Robert M{\"u}ller}.} \bibinfo{year}{2019}\natexlab{}.
\newblock \showarticletitle{Unmasking Clever Hans predictors and assessing what machines really learn}.
\newblock \bibinfo{journal}{\emph{Nature Communications}} \bibinfo{volume}{10}, \bibinfo{number}{1} (\bibinfo{date}{11 Mar} \bibinfo{year}{2019}), \bibinfo{pages}{1096}.
\newblock
\showISSN{2041-1723}
\href{https://doi.org/10.1038/s41467-019-08987-4}{doi:\nolinkurl{10.1038/s41467-019-08987-4}}


\bibitem[Leon et~al\mbox{.}(2017)]%
        {Leon:2017:EvaluatingTheEffectOfVotingMethodsOnEnsembleBasedClassification}
\bibfield{author}{\bibinfo{person}{Florin Leon}, \bibinfo{person}{Sabina-Adriana Floria}, {and} \bibinfo{person}{Costin Bădică}.} \bibinfo{year}{2017}\natexlab{}.
\newblock \showarticletitle{Evaluating the effect of voting methods on ensemble-based classification}. In \bibinfo{booktitle}{\emph{2017 IEEE International Conference on INnovations in Intelligent SysTems and Applications (INISTA)}}. \bibinfo{pages}{1--6}.
\newblock
\href{https://doi.org/10.1109/INISTA.2017.8001122}{doi:\nolinkurl{10.1109/INISTA.2017.8001122}}


\bibitem[Li et~al\mbox{.}(2016)]%
        {Li:2017:UnderstandingNeuralNetworksthroughRepresentationErasure}
\bibfield{author}{\bibinfo{person}{Jiwei Li}, \bibinfo{person}{Will Monroe}, {and} \bibinfo{person}{Dan Jurafsky}.} \bibinfo{year}{2016}\natexlab{}.
\newblock \showarticletitle{Understanding Neural Networks through Representation Erasure.}
\newblock \bibinfo{journal}{\emph{CoRR}}  \bibinfo{volume}{abs/1612.08220} (\bibinfo{year}{2016}).
\newblock
\urldef\tempurl%
\url{http://dblp.uni-trier.de/db/journals/corr/corr1612.html#LiMJ16a}
\showURL{%
\tempurl}


\bibitem[Liao and Wortman(2024)]%
        {Liao:2023:AITransparencyintheAgeofLLMsAHumanCenteredResearchRoadmap}
\bibfield{author}{\bibinfo{person}{Q.~Vera Liao} {and} \bibinfo{person}{Vaughan~Jennifer Wortman}.} \bibinfo{year}{2024}\natexlab{}.
\newblock \showarticletitle{{AI} {Transparency} in the {Age} of {LLMs}: A {Human}-{Centered} {Research} {Roadmap}}.
\newblock \bibinfo{journal}{\emph{Harvard Data Science Review}} \bibinfo{number}{Special Issue 5} (\bibinfo{date}{feb 29} \bibinfo{year}{2024}).
\newblock
\urldef\tempurl%
\url{https://hdsr.mitpress.mit.edu/pub/aelql9qy}
\showURL{%
\tempurl}


\bibitem[Linhardt et~al\mbox{.}(2024)]%
        {Linhardt:2024:PreemptivelyPruningCleverHansStrategiesinDeepNeuralNetworks}
\bibfield{author}{\bibinfo{person}{Lorenz Linhardt}, \bibinfo{person}{Klaus-Robert Müller}, {and} \bibinfo{person}{Grégoire Montavon}.} \bibinfo{year}{2024}\natexlab{}.
\newblock \showarticletitle{Preemptively pruning Clever-Hans strategies in deep neural networks}.
\newblock \bibinfo{journal}{\emph{Information Fusion}}  \bibinfo{volume}{103} (\bibinfo{year}{2024}), \bibinfo{pages}{102094}.
\newblock
\showISSN{1566-2535}
\href{https://doi.org/10.1016/j.inffus.2023.102094}{doi:\nolinkurl{10.1016/j.inffus.2023.102094}}


\bibitem[Ma et~al\mbox{.}(2018)]%
        {Ma:2018:DeepGauge:MultiGranularityTestingCriteriaForDeepLearningSystems}
\bibfield{author}{\bibinfo{person}{Lei Ma}, \bibinfo{person}{Felix Juefei-Xu}, \bibinfo{person}{Fuyuan Zhang}, \bibinfo{person}{Jiyuan Sun}, \bibinfo{person}{Minhui Xue}, \bibinfo{person}{Bo Li}, \bibinfo{person}{Chunyang Chen}, \bibinfo{person}{Ting Su}, \bibinfo{person}{Li Li}, \bibinfo{person}{Yang Liu}, \bibinfo{person}{Jianjun Zhao}, {and} \bibinfo{person}{Yadong Wang}.} \bibinfo{year}{2018}\natexlab{}.
\newblock \showarticletitle{DeepGauge: multi-granularity testing criteria for deep learning systems}. In \bibinfo{booktitle}{\emph{The 33rd ACM/IEEE International Conference on Automated Software Engineering}} (Montpellier, France) \emph{(\bibinfo{series}{ASE '18})}. \bibinfo{publisher}{Association for Computing Machinery}, \bibinfo{address}{New York, USA}, \bibinfo{pages}{120–131}.
\newblock
\showISBNx{9781450359375}
\href{https://doi.org/10.1145/3238147.3238202}{doi:\nolinkurl{10.1145/3238147.3238202}}


\bibitem[Mariotti et~al\mbox{.}(2024)]%
        {Mariotti:2024:TextFocus:AssessingtheFaithfulnessofFeatureAttributionMethodsExplanationsinNaturalLanguageProcessing}
\bibfield{author}{\bibinfo{person}{Ettore Mariotti}, \bibinfo{person}{Anna Arias-Duart}, \bibinfo{person}{Michele Cafagna}, \bibinfo{person}{Albert Gatt}, \bibinfo{person}{Dario Garcia-Gasulla}, {and} \bibinfo{person}{Jose~Maria Alonso-Moral}.} \bibinfo{year}{2024}\natexlab{}.
\newblock \showarticletitle{TextFocus: Assessing the Faithfulness of Feature Attribution Methods Explanations in Natural Language Processing}.
\newblock \bibinfo{journal}{\emph{IEEE Access}} (\bibinfo{year}{2024}), \bibinfo{pages}{1--1}.
\newblock
\href{https://doi.org/10.1109/ACCESS.2024.3408062}{doi:\nolinkurl{10.1109/ACCESS.2024.3408062}}


\bibitem[Mart\'{\i}nez et~al\mbox{.}(2022)]%
        {Silverio:2022:SoftwareEngineeringforAIBasedSystemsASurvey}
\bibfield{author}{\bibinfo{person}{Fern\'{a}ndez~Silverio Mart\'{\i}nez}, \bibinfo{person}{Justus Bogner}, \bibinfo{person}{Xavier Franch}, \bibinfo{person}{Marc Oriol}, \bibinfo{person}{Julien Siebert}, \bibinfo{person}{Adam Trendowicz}, \bibinfo{person}{Anna~Maria Vollmer}, {and} \bibinfo{person}{Stefan Wagner}.} \bibinfo{year}{2022}\natexlab{}.
\newblock \showarticletitle{Software Engineering for AI-Based Systems: A Survey}.
\newblock \bibinfo{journal}{\emph{ACM Trans. Softw. Eng. Methodol.}} \bibinfo{volume}{31}, \bibinfo{number}{2}, Article \bibinfo{articleno}{37e} (\bibinfo{date}{April} \bibinfo{year}{2022}), \bibinfo{numpages}{59}~pages.
\newblock
\showISSN{1049-331X}
\href{https://doi.org/10.1145/3487043}{doi:\nolinkurl{10.1145/3487043}}


\bibitem[Mathew et~al\mbox{.}(2021)]%
        {Mathew:2021:HateXplain:ABenchmarkDatasetforExplainableHateSpeechDetection}
\bibfield{author}{\bibinfo{person}{Binny Mathew}, \bibinfo{person}{Punyajoy Saha}, \bibinfo{person}{Seid~Muhie Yimam}, \bibinfo{person}{Chris Biemann}, \bibinfo{person}{Pawan Goyal}, {and} \bibinfo{person}{Animesh Mukherjee}.} \bibinfo{year}{2021}\natexlab{}.
\newblock \showarticletitle{HateXplain: A Benchmark Dataset for Explainable Hate Speech Detection}.
\newblock \bibinfo{journal}{\emph{Proceedings of the AAAI Conference on Artificial Intelligence}} \bibinfo{volume}{35}, \bibinfo{number}{17}, \bibinfo{pages}{14867--14875}.
\newblock
\href{https://doi.org/10.1609/aaai.v35i17.17745}{doi:\nolinkurl{10.1609/aaai.v35i17.17745}}


\bibitem[Miller(1995)]%
        {Miller:1995:WordNetALexicalDatabaseforEnglish}
\bibfield{author}{\bibinfo{person}{George~A. Miller}.} \bibinfo{year}{1995}\natexlab{}.
\newblock \showarticletitle{WordNet: a lexical database for English}.
\newblock \bibinfo{journal}{\emph{Commun. ACM}} \bibinfo{volume}{38}, \bibinfo{number}{11} (\bibinfo{date}{nov} \bibinfo{year}{1995}), \bibinfo{pages}{39–41}.
\newblock
\showISSN{0001-0782}
\href{https://doi.org/10.1145/219717.219748}{doi:\nolinkurl{10.1145/219717.219748}}


\bibitem[Mohebbi et~al\mbox{.}(2021)]%
        {Mohebbi:2021:ExploringtheRoleofBERTTokenRepresentationstoExplainSentenceProbingResults}
\bibfield{author}{\bibinfo{person}{Hosein Mohebbi}, \bibinfo{person}{Ali Modarressi}, {and} \bibinfo{person}{Mohammad~Taher Pilehvar}.} \bibinfo{year}{2021}\natexlab{}.
\newblock \showarticletitle{Exploring the Role of {BERT} Token Representations to Explain Sentence Probing Results}. In \bibinfo{booktitle}{\emph{Empirical Methods in Natural Language Processing}}. \bibinfo{publisher}{Association for Computational Linguistics}, \bibinfo{address}{Dominican Republic}, \bibinfo{pages}{792--806}.
\newblock
\href{https://doi.org/10.18653/v1/2021.emnlp-main.61}{doi:\nolinkurl{10.18653/v1/2021.emnlp-main.61}}


\bibitem[Morris and Hirst(2004)]%
        {Morris:2004:NonClassicalLexicalSemanticRelations}
\bibfield{author}{\bibinfo{person}{Jane Morris} {and} \bibinfo{person}{Graeme Hirst}.} \bibinfo{year}{2004}\natexlab{}.
\newblock \showarticletitle{Non-Classical Lexical Semantic Relations}. In \bibinfo{booktitle}{\emph{The Computational Lexical Semantics Workshop}}. \bibinfo{publisher}{Association for Computational Linguistics}, \bibinfo{address}{Boston, USA}, \bibinfo{pages}{46--51}.
\newblock
\urldef\tempurl%
\url{https://aclanthology.org/W04-2607}
\showURL{%
\tempurl}


\bibitem[Morris et~al\mbox{.}(2020)]%
        {Morris:2020:TextAttack:AFrameworkforAdversarialAttacksDataAugmentationandAdversarialTraininginNLP}
\bibfield{author}{\bibinfo{person}{John Morris}, \bibinfo{person}{Eli Lifland}, \bibinfo{person}{Jin~Yong Yoo}, \bibinfo{person}{Jake Grigsby}, \bibinfo{person}{Di Jin}, {and} \bibinfo{person}{Yanjun Qi}.} \bibinfo{year}{2020}\natexlab{}.
\newblock \showarticletitle{{T}ext{A}ttack: A Framework for Adversarial Attacks, Data Augmentation, and Adversarial Training in {NLP}}. In \bibinfo{booktitle}{\emph{Empirical Methods in Natural Language Processing}}. \bibinfo{publisher}{Association for Computational Linguistics}, \bibinfo{pages}{119--126}.
\newblock
\href{https://doi.org/10.18653/v1/2020.emnlp-demos.16}{doi:\nolinkurl{10.18653/v1/2020.emnlp-demos.16}}


\bibitem[Mrk{\v{s}}i{\'c} et~al\mbox{.}(2016)]%
        {Mrksic:2016:CounterfittingWordVectorstoLinguisticConstraints}
\bibfield{author}{\bibinfo{person}{Nikola Mrk{\v{s}}i{\'c}}, \bibinfo{person}{Diarmuid {\'O}~S{\'e}aghdha}, \bibinfo{person}{Blaise Thomson}, \bibinfo{person}{Milica Ga{\v{s}}i{\'c}}, \bibinfo{person}{Lina~M. Rojas-Barahona}, \bibinfo{person}{Pei-Hao Su}, \bibinfo{person}{David Vandyke}, \bibinfo{person}{Tsung-Hsien Wen}, {and} \bibinfo{person}{Steve Young}.} \bibinfo{year}{2016}\natexlab{}.
\newblock \showarticletitle{Counter-fitting Word Vectors to Linguistic Constraints}. In \bibinfo{booktitle}{\emph{The North {A}merican Chapter of the Association for Computational Linguistics: Human Language Technologies}}. \bibinfo{address}{San Diego, California}, \bibinfo{pages}{142--148}.
\newblock
\href{https://doi.org/10.18653/v1/N16-1018}{doi:\nolinkurl{10.18653/v1/N16-1018}}


\bibitem[Nguyen et~al\mbox{.}(2015)]%
        {Anh:2015:DeepNeuralNetworksAreEasilyFooledHighConfidencePredictionsForUnrecognizableImages}
\bibfield{author}{\bibinfo{person}{Anh Nguyen}, \bibinfo{person}{Jason Yosinski}, {and} \bibinfo{person}{Jeff Clune}.} \bibinfo{year}{2015}\natexlab{}.
\newblock \showarticletitle{Deep neural networks are easily fooled: High confidence predictions for unrecognizable images}. In \bibinfo{booktitle}{\emph{2015 IEEE Conference on Computer Vision and Pattern Recognition (CVPR)}}. \bibinfo{pages}{427--436}.
\newblock
\href{https://doi.org/10.1109/CVPR.2015.7298640}{doi:\nolinkurl{10.1109/CVPR.2015.7298640}}


\bibitem[Nielsen(2016)]%
        {Nielsen:2016:HierarchicalClustering}
\bibfield{author}{\bibinfo{person}{Frank Nielsen}.} \bibinfo{year}{2016}\natexlab{}.
\newblock \bibinfo{booktitle}{\emph{Hierarchical Clustering}}.
\newblock \bibinfo{publisher}{Springer International Publishing}, \bibinfo{address}{Cham}, \bibinfo{pages}{195--211}.
\newblock
\showISBNx{978-3-319-21903-5}
\href{https://doi.org/10.1007/978-3-319-21903-5_8}{doi:\nolinkurl{10.1007/978-3-319-21903-5_8}}


\bibitem[Ortu et~al\mbox{.}(2015)]%
        {Ortu:2015:AreBulliesMoreProductiveEmpiricalStudyofAffectivenessvsIssueFixingTime}
\bibfield{author}{\bibinfo{person}{Marco Ortu}, \bibinfo{person}{Bram Adams}, \bibinfo{person}{Giuseppe Destefanis}, \bibinfo{person}{Parastou Tourani}, \bibinfo{person}{Michele Marchesi}, {and} \bibinfo{person}{Roberto Tonelli}.} \bibinfo{year}{2015}\natexlab{}.
\newblock \showarticletitle{Are Bullies More Productive? Empirical Study of Affectiveness vs. Issue Fixing Time}. In \bibinfo{booktitle}{\emph{2015 IEEE/ACM 12th Working Conference on Mining Software Repositories}}. \bibinfo{pages}{303--313}.
\newblock
\href{https://doi.org/10.1109/MSR.2015.35}{doi:\nolinkurl{10.1109/MSR.2015.35}}


\bibitem[Ovadia et~al\mbox{.}(2019)]%
        {Ovadia:2019:CanYouTrustYourModelsUncertaintyEvaluatingPredictiveUncertaintyUnderDatasetShift}
\bibfield{author}{\bibinfo{person}{Yaniv Ovadia}, \bibinfo{person}{Emily Fertig}, \bibinfo{person}{Jie Ren}, \bibinfo{person}{Zachary Nado}, \bibinfo{person}{D. Sculley}, \bibinfo{person}{Sebastian Nowozin}, \bibinfo{person}{Joshua Dillon}, \bibinfo{person}{Balaji Lakshminarayanan}, {and} \bibinfo{person}{Jasper Snoek}.} \bibinfo{year}{2019}\natexlab{}.
\newblock \showarticletitle{Can you trust your model\textquotesingle s uncertainty? Evaluating predictive uncertainty under dataset shift}. In \bibinfo{booktitle}{\emph{Advances in Neural Information Processing Systems}}, Vol.~\bibinfo{volume}{32}. \bibinfo{publisher}{Curran Associates}.
\newblock
\urldef\tempurl%
\url{https://proceedings.neurips.cc/paper_files/paper/2019/file/8558cb408c1d76621371888657d2eb1d-Paper.pdf}
\showURL{%
\tempurl}


\bibitem[Panichella et~al\mbox{.}(2015)]%
        {Panichella:2015:HowCanIImproveMyAppClassifyingUserReviewsForSoftwareMaintenanceAndEvolution}
\bibfield{author}{\bibinfo{person}{Sebastiano Panichella}, \bibinfo{person}{Andrea Di~Sorbo}, \bibinfo{person}{Emitza Guzman}, \bibinfo{person}{Corrado~A. Visaggio}, \bibinfo{person}{Gerardo Canfora}, {and} \bibinfo{person}{Harald~C. Gall}.} \bibinfo{year}{2015}\natexlab{}.
\newblock \showarticletitle{How can i improve my app? Classifying user reviews for software maintenance and evolution}. In \bibinfo{booktitle}{\emph{IEEE International Conference on Software Maintenance and Evolution}}. \bibinfo{pages}{281--290}.
\newblock
\href{https://doi.org/10.1109/ICSM.2015.7332474}{doi:\nolinkurl{10.1109/ICSM.2015.7332474}}


\bibitem[Pei et~al\mbox{.}(2017)]%
        {Pei:2017:DeepXplore:AutomatedWhiteboxTestingofDeepLearningSystems}
\bibfield{author}{\bibinfo{person}{Kexin Pei}, \bibinfo{person}{Yinzhi Cao}, \bibinfo{person}{Junfeng Yang}, {and} \bibinfo{person}{Suman Jana}.} \bibinfo{year}{2017}\natexlab{}.
\newblock \showarticletitle{DeepXplore: Automated Whitebox Testing of Deep Learning Systems}. In \bibinfo{booktitle}{\emph{26th Symposium on Operating Systems Principles}} (Shanghai, China) \emph{(\bibinfo{series}{SOSP '17})}. \bibinfo{publisher}{Association for Computing Machinery}, \bibinfo{address}{New York, USA}, \bibinfo{pages}{1–18}.
\newblock
\showISBNx{9781450350853}
\href{https://doi.org/10.1145/3132747.3132785}{doi:\nolinkurl{10.1145/3132747.3132785}}


\bibitem[Pennington et~al\mbox{.}(2014)]%
        {Pennington:2014:Glove:GlobalVectorsforWordRepresentation}
\bibfield{author}{\bibinfo{person}{Jeffrey Pennington}, \bibinfo{person}{Richard Socher}, {and} \bibinfo{person}{Christopher Manning}.} \bibinfo{year}{2014}\natexlab{}.
\newblock \showarticletitle{{G}lo{V}e: Global Vectors for Word Representation}. In \bibinfo{booktitle}{\emph{The 2014 Conference on Empirical Methods in Natural Language Processing ({EMNLP})}}. \bibinfo{publisher}{Association for Computational Linguistics}, \bibinfo{address}{Doha, Qatar}, \bibinfo{pages}{1532--1543}.
\newblock
\href{https://doi.org/10.3115/v1/D14-1162}{doi:\nolinkurl{10.3115/v1/D14-1162}}


\bibitem[Pérez-Verdejo et~al\mbox{.}(2020)]%
        {Pérez:2020:ASystematicLiteratureReviewonMachineLearningforAutomatedRequirementsClassification}
\bibfield{author}{\bibinfo{person}{J.~Manuel Pérez-Verdejo}, \bibinfo{person}{Angel~J. Sánchez-García}, {and} \bibinfo{person}{Jorge~Octavio Ocharán-Hernández}.} \bibinfo{year}{2020}\natexlab{}.
\newblock \showarticletitle{A Systematic Literature Review on Machine Learning for Automated Requirements Classification}. In \bibinfo{booktitle}{\emph{2020 8th International Conference in Software Engineering Research and Innovation (CONISOFT)}}. \bibinfo{pages}{21--28}.
\newblock
\href{https://doi.org/10.1109/CONISOFT50191.2020.00014}{doi:\nolinkurl{10.1109/CONISOFT50191.2020.00014}}


\bibitem[Rahman et~al\mbox{.}(2015)]%
        {Rahman:2015:RecommendingInsightfulCommentsForSourceCodeUsingCrowdsourcedKnowledge}
\bibfield{author}{\bibinfo{person}{Mohammad~Masudur Rahman}, \bibinfo{person}{Chanchal~K. Roy}, {and} \bibinfo{person}{Iman Keivanloo}.} \bibinfo{year}{2015}\natexlab{}.
\newblock \showarticletitle{Recommending insightful comments for source code using crowdsourced knowledge}. In \bibinfo{booktitle}{\emph{2015 IEEE 15th International Working Conference on Source Code Analysis and Manipulation (SCAM)}}. \bibinfo{pages}{81--90}.
\newblock
\href{https://doi.org/10.1109/SCAM.2015.7335404}{doi:\nolinkurl{10.1109/SCAM.2015.7335404}}


\bibitem[Ramanagopal et~al\mbox{.}(2018)]%
        {Ramanagopal:2018:FailingtoLearn:AutonomouslyIdentifyingPerceptionFailuresforSelfDrivingCars}
\bibfield{author}{\bibinfo{person}{Manikandasriram~Srinivasan Ramanagopal}, \bibinfo{person}{Cyrus Anderson}, \bibinfo{person}{Ram Vasudevan}, {and} \bibinfo{person}{Matthew Johnson-Roberson}.} \bibinfo{year}{2018}\natexlab{}.
\newblock \showarticletitle{Failing to Learn: Autonomously Identifying Perception Failures for Self-Driving Cars}.
\newblock \bibinfo{journal}{\emph{IEEE Robotics and Automation Letters}} \bibinfo{volume}{3}, \bibinfo{number}{4} (\bibinfo{year}{2018}), \bibinfo{pages}{3860--3867}.
\newblock
\href{https://doi.org/10.1109/LRA.2018.2857402}{doi:\nolinkurl{10.1109/LRA.2018.2857402}}


\bibitem[Rechkemmer and Yin(2022)]%
        {Rechkemmer:2022:WhenConfidenceMeetsAccuracyExploringtheEffectsofMultiplePerformanceIndicatorsonTrustinMachineLearningModels}
\bibfield{author}{\bibinfo{person}{Amy Rechkemmer} {and} \bibinfo{person}{Ming Yin}.} \bibinfo{year}{2022}\natexlab{}.
\newblock \showarticletitle{When Confidence Meets Accuracy: Exploring the Effects of Multiple Performance Indicators on Trust in Machine Learning Models}. In \bibinfo{booktitle}{\emph{Human Factors in Computing Systems}} (LA, USA) \emph{(\bibinfo{series}{CHI '22})}. \bibinfo{publisher}{Association for Computing Machinery}, \bibinfo{address}{NY, USA}, Article \bibinfo{articleno}{535}, \bibinfo{numpages}{14}~pages.
\newblock
\showISBNx{9781450391573}
\href{https://doi.org/10.1145/3491102.3501967}{doi:\nolinkurl{10.1145/3491102.3501967}}


\bibitem[Reimers and Gurevych(2019)]%
        {Reimers:2019:SentenceBERT:SentenceEmbeddingsusingSiameseBERTNetworks}
\bibfield{author}{\bibinfo{person}{Nils Reimers} {and} \bibinfo{person}{Iryna Gurevych}.} \bibinfo{year}{2019}\natexlab{}.
\newblock \showarticletitle{Sentence-{BERT}: Sentence Embeddings using {S}iamese {BERT}-Networks}. In \bibinfo{booktitle}{\emph{Empirical Methods in Natural Language Processing and International Joint Conference on Natural Language Processing}}. \bibinfo{publisher}{Association for Computational Linguistics}, \bibinfo{address}{Hong Kong, China}, \bibinfo{pages}{3982--3992}.
\newblock
\href{https://doi.org/10.18653/v1/D19-1410}{doi:\nolinkurl{10.18653/v1/D19-1410}}


\bibitem[Ribeiro et~al\mbox{.}(2016)]%
        {Ribeiro:2016:WhyShouldITrustYouExplainingthePredictionsofAnyClassifier}
\bibfield{author}{\bibinfo{person}{Marco~Tulio Ribeiro}, \bibinfo{person}{Sameer Singh}, {and} \bibinfo{person}{Carlos Guestrin}.} \bibinfo{year}{2016}\natexlab{}.
\newblock \showarticletitle{"Why Should I Trust You?": Explaining the Predictions of Any Classifier}. In \bibinfo{booktitle}{\emph{The 22nd ACM SIGKDD International Conference on Knowledge Discovery and Data Mining}} (San Francisco, California, USA) \emph{(\bibinfo{series}{KDD '16})}. \bibinfo{pages}{1135–1144}.
\newblock
\showISBNx{9781450342322}
\href{https://doi.org/10.1145/2939672.2939778}{doi:\nolinkurl{10.1145/2939672.2939778}}


\bibitem[Riccio et~al\mbox{.}(2020)]%
        {Riccio:2020:TestingMachineLearningBasedSystems:ASystematicMapping}
\bibfield{author}{\bibinfo{person}{Vincenzo Riccio}, \bibinfo{person}{Gunel Jahangirova}, \bibinfo{person}{Andrea Stocco}, \bibinfo{person}{Nargiz Humbatova}, \bibinfo{person}{Michael Weiss}, {and} \bibinfo{person}{Paolo Tonella}.} \bibinfo{year}{2020}\natexlab{}.
\newblock \showarticletitle{Testing machine learning based systems: a systematic mapping}.
\newblock \bibinfo{journal}{\emph{Empirical Software Engineering}} \bibinfo{volume}{25}, \bibinfo{number}{6} (\bibinfo{date}{01 Nov} \bibinfo{year}{2020}), \bibinfo{pages}{5193--5254}.
\newblock
\showISSN{1573-7616}
\href{https://doi.org/10.1007/s10664-020-09881-0}{doi:\nolinkurl{10.1007/s10664-020-09881-0}}


\bibitem[Ross et~al\mbox{.}(2017)]%
        {Andrew:2017:RightfortheRightReasonsTrainingDifferentiableModelsbyConstrainingtheirExplanations}
\bibfield{author}{\bibinfo{person}{Andrew~Slavin Ross}, \bibinfo{person}{Michael~C. Hughes}, {and} \bibinfo{person}{Finale Doshi-Velez}.} \bibinfo{year}{2017}\natexlab{}.
\newblock \showarticletitle{Right for the Right Reasons: Training Differentiable Models by Constraining their Explanations}. In \bibinfo{booktitle}{\emph{Joint Conference on Artificial Intelligence}}. \bibinfo{pages}{2662--2670}.
\newblock
\href{https://doi.org/10.24963/ijcai.2017/371}{doi:\nolinkurl{10.24963/ijcai.2017/371}}


\bibitem[Schlegel et~al\mbox{.}(2022)]%
        {Schlegel:2022:TowardsHumanCentredExplainabilityBenchmarksForTextClassification}
\bibfield{author}{\bibinfo{person}{Viktor Schlegel}, \bibinfo{person}{Erick Mendez~Guzman}, {and} \bibinfo{person}{Riza Batista-Navarro}.} \bibinfo{year}{2022}\natexlab{}.
\newblock \showarticletitle{Towards Human-Centred Explainability Benchmarks For Text Classification}. In \bibinfo{booktitle}{\emph{16th International AAAI Conference on Web and Social Media}}. \bibinfo{publisher}{AAAI}, \bibinfo{address}{USA}, \bibinfo{pages}{1}.
\newblock


\bibitem[Schramowski et~al\mbox{.}(2020)]%
        {Schramowski:2020:MakingDeepNeuralNetworksRightForTheRightScientificReasonsByInteractingWithTheirExplanations}
\bibfield{author}{\bibinfo{person}{Patrick Schramowski}, \bibinfo{person}{Wolfgang Stammer}, \bibinfo{person}{Stefano Teso}, \bibinfo{person}{Anna Brugger}, \bibinfo{person}{Franziska Herbert}, \bibinfo{person}{Xiaoting Shao}, \bibinfo{person}{Hans-Georg Luigs}, \bibinfo{person}{Anne-Katrin Mahlein}, {and} \bibinfo{person}{Kristian Kersting}.} \bibinfo{year}{2020}\natexlab{}.
\newblock \showarticletitle{Making deep neural networks right for the right scientific reasons by interacting with their explanations}.
\newblock \bibinfo{journal}{\emph{Nature Machine Intelligence}} \bibinfo{volume}{2}, \bibinfo{number}{8} (\bibinfo{date}{Aug} \bibinfo{year}{2020}), \bibinfo{pages}{476--486}.
\newblock
\showISSN{2522-5839}
\href{https://doi.org/10.1038/s42256-020-0212-3}{doi:\nolinkurl{10.1038/s42256-020-0212-3}}


\bibitem[Schulte et~al\mbox{.}(2024)]%
        {Schulte:2024:StudyingTheExplanationsForTheAutomatedPredictionOfBugAndNonbugIssuesUsingLimeAndShap}
\bibfield{author}{\bibinfo{person}{Lukas Schulte}, \bibinfo{person}{Benjamin Ledel}, {and} \bibinfo{person}{Steffen Herbold}.} \bibinfo{year}{2024}\natexlab{}.
\newblock \showarticletitle{Studying the explanations for the automated prediction of bug and non-bug issues using LIME and SHAP}.
\newblock \bibinfo{journal}{\emph{Empirical Software Engineering}} \bibinfo{volume}{29}, \bibinfo{number}{4} (\bibinfo{date}{13 Jun} \bibinfo{year}{2024}), \bibinfo{pages}{93}.
\newblock
\showISSN{1573-7616}
\href{https://doi.org/10.1007/s10664-024-10469-1}{doi:\nolinkurl{10.1007/s10664-024-10469-1}}


\bibitem[Shazeer et~al\mbox{.}(2016)]%
        {Noam:2016:Swivel:ImprovingEmbeddingsbyNoticingWhatsMissing}
\bibfield{author}{\bibinfo{person}{Noam Shazeer}, \bibinfo{person}{Ryan Doherty}, \bibinfo{person}{Colin Evans}, {and} \bibinfo{person}{Chris Waterson}.} \bibinfo{year}{2016}\natexlab{}.
\newblock \showarticletitle{Swivel: Improving Embeddings by Noticing What's Missing}.
\newblock \bibinfo{journal}{\emph{CoRR}}  \bibinfo{volume}{abs/1602.02215} (\bibinfo{year}{2016}), \bibinfo{numpages}{9}~pages.
\newblock
\showeprint[arXiv]{1602.02215}
\urldef\tempurl%
\url{http://arxiv.org/abs/1602.02215}
\showURL{%
\tempurl}


\bibitem[Sinha et~al\mbox{.}(2016)]%
        {Sinha:2016:AnalyzingDeveloperSentimentInCommitLogs}
\bibfield{author}{\bibinfo{person}{Vinayak Sinha}, \bibinfo{person}{Alina Lazar}, {and} \bibinfo{person}{Bonita Sharif}.} \bibinfo{year}{2016}\natexlab{}.
\newblock \showarticletitle{Analyzing developer sentiment in commit logs}. In \bibinfo{booktitle}{\emph{The 13th International Conference on Mining Software Repositories}} (Austin, Texas) \emph{(\bibinfo{series}{MSR '16})}. \bibinfo{publisher}{Association for Computing Machinery}, \bibinfo{address}{New York, NY, USA}, \bibinfo{pages}{520–523}.
\newblock
\showISBNx{9781450341868}
\href{https://doi.org/10.1145/2901739.2903501}{doi:\nolinkurl{10.1145/2901739.2903501}}


\bibitem[Slimani(2013)]%
        {Thabet:2013:DescriptionandEvaluationofSemanticSimilarityMeasuresApproaches}
\bibfield{author}{\bibinfo{person}{Thabet Slimani}.} \bibinfo{year}{2013}\natexlab{}.
\newblock \showarticletitle{Description and Evaluation of Semantic Similarity Measures Approaches}.
\newblock \bibinfo{journal}{\emph{International Journal of Computer Applications}} \bibinfo{volume}{80}, \bibinfo{number}{10} (\bibinfo{date}{October} \bibinfo{year}{2013}), \bibinfo{pages}{25--33}.
\newblock
\showISSN{0975-8887}
\href{https://doi.org/10.5120/13897-1851}{doi:\nolinkurl{10.5120/13897-1851}}


\bibitem[Thomas et~al\mbox{.}(2019)]%
        {Thomas:2019:AnalyzingNeuroimagingDataThroughRecurrentDeepLearningModels}
\bibfield{author}{\bibinfo{person}{Armin~W. Thomas}, \bibinfo{person}{Hauke~R. Heekeren}, \bibinfo{person}{Klaus-Robert Müller}, {and} \bibinfo{person}{Wojciech Samek}.} \bibinfo{year}{2019}\natexlab{}.
\newblock \showarticletitle{Analyzing Neuroimaging Data Through Recurrent Deep Learning Models}.
\newblock \bibinfo{journal}{\emph{Frontiers in Neuroscience}}  \bibinfo{volume}{13} (\bibinfo{year}{2019}).
\newblock
\showISSN{1662-453X}
\href{https://doi.org/10.3389/fnins.2019.01321}{doi:\nolinkurl{10.3389/fnins.2019.01321}}


\bibitem[Tian et~al\mbox{.}(2018)]%
        {Tian:2018:DeepTest:AutomatedTestingOfDeepNeuralNetworkDrivenAutonomousCars}
\bibfield{author}{\bibinfo{person}{Yuchi Tian}, \bibinfo{person}{Kexin Pei}, \bibinfo{person}{Suman Jana}, {and} \bibinfo{person}{Baishakhi Ray}.} \bibinfo{year}{2018}\natexlab{}.
\newblock \showarticletitle{DeepTest: automated testing of deep-neural-network-driven autonomous cars}. In \bibinfo{booktitle}{\emph{The 40th International Conference on Software Engineering}} (Sweden) \emph{(\bibinfo{series}{ICSE '18})}. \bibinfo{publisher}{Association for Computing Machinery}, \bibinfo{address}{USA}, \bibinfo{pages}{303–314}.
\newblock
\showISBNx{9781450356381}
\href{https://doi.org/10.1145/3180155.3180220}{doi:\nolinkurl{10.1145/3180155.3180220}}


\bibitem[Torregrossa et~al\mbox{.}(2021)]%
        {Torregrossa:2021:ASurveyOnTrainingAndEvaluationOfWordEmbeddings}
\bibfield{author}{\bibinfo{person}{Fran{\c{c}}ois Torregrossa}, \bibinfo{person}{Robin Allesiardo}, \bibinfo{person}{Vincent Claveau}, \bibinfo{person}{Nihel Kooli}, {and} \bibinfo{person}{Guillaume Gravier}.} \bibinfo{year}{2021}\natexlab{}.
\newblock \showarticletitle{A survey on training and evaluation of word embeddings}.
\newblock \bibinfo{journal}{\emph{International Journal of Data Science and Analytics}} \bibinfo{volume}{11}, \bibinfo{number}{2} (\bibinfo{date}{01 Mar} \bibinfo{year}{2021}), \bibinfo{pages}{85--103}.
\newblock
\showISSN{2364-4168}
\href{https://doi.org/10.1007/s41060-021-00242-8}{doi:\nolinkurl{10.1007/s41060-021-00242-8}}


\bibitem[Treviso et~al\mbox{.}(2023)]%
        {Treviso:2023:CREST:AJointFrameworkforRationalizationandCounterfactualTextGeneration}
\bibfield{author}{\bibinfo{person}{Marcos Treviso}, \bibinfo{person}{Alexis Ross}, \bibinfo{person}{Nuno~M. Guerreiro}, {and} \bibinfo{person}{Andr{\'e} Martins}.} \bibinfo{year}{2023}\natexlab{}.
\newblock \showarticletitle{{CREST}: A Joint Framework for Rationalization and Counterfactual Text Generation}. In \bibinfo{booktitle}{\emph{The 61st Annual Meeting of the Association for Computational Linguistics (Volume 1: Long Papers)}}. \bibinfo{address}{Toronto, Canada}, \bibinfo{pages}{15109--15126}.
\newblock
\href{https://doi.org/10.18653/v1/2023.acl-long.842}{doi:\nolinkurl{10.18653/v1/2023.acl-long.842}}


\bibitem[Wiegreffe and Pinter(2019)]%
        {Wiegreffe:2019:AttentionIsNotNotExplanation}
\bibfield{author}{\bibinfo{person}{Sarah Wiegreffe} {and} \bibinfo{person}{Yuval Pinter}.} \bibinfo{year}{2019}\natexlab{}.
\newblock \showarticletitle{Attention is not not Explanation}. In \bibinfo{booktitle}{\emph{The 2019 Conference on Empirical Methods in Natural Language Processing and the 9th International Joint Conference on Natural Language Processing (EMNLP-IJCNLP)}}. \bibinfo{publisher}{Association for Computational Linguistics}, \bibinfo{address}{Hong Kong, China}, \bibinfo{pages}{11--20}.
\newblock
\href{https://doi.org/10.18653/v1/D19-1002}{doi:\nolinkurl{10.18653/v1/D19-1002}}


\bibitem[Xie et~al\mbox{.}(2020)]%
        {Xie:2020:METTLE:AMETamorphicTestingApproachtoAssessingandValidatingUnsupervisedMachineLearningSystems}
\bibfield{author}{\bibinfo{person}{Xiaoyuan Xie}, \bibinfo{person}{Zhiyi Zhang}, \bibinfo{person}{Tsong~Yueh Chen}, \bibinfo{person}{Yang Liu}, \bibinfo{person}{Pak-Lok Poon}, {and} \bibinfo{person}{Baowen Xu}.} \bibinfo{year}{2020}\natexlab{}.
\newblock \showarticletitle{METTLE: A METamorphic Testing Approach to Assessing and Validating Unsupervised Machine Learning Systems}.
\newblock \bibinfo{journal}{\emph{IEEE Transactions on Reliability}} \bibinfo{volume}{69}, \bibinfo{number}{4} (\bibinfo{year}{2020}), \bibinfo{pages}{1293--1322}.
\newblock
\href{https://doi.org/10.1109/TR.2020.2972266}{doi:\nolinkurl{10.1109/TR.2020.2972266}}


\bibitem[Ye et~al\mbox{.}(2024)]%
        {Ye:2024:SpuriousCorrelationsinMachineLearningASurvey}
\bibfield{author}{\bibinfo{person}{Wenqian Ye}, \bibinfo{person}{Guangtao Zheng}, \bibinfo{person}{Xu Cao}, \bibinfo{person}{Yunsheng Ma}, {and} \bibinfo{person}{Aidong Zhang}.} \bibinfo{year}{2024}\natexlab{}.
\newblock \bibinfo{title}{Spurious Correlations in Machine Learning: A Survey}.
\newblock
\showeprint[arxiv]{2402.12715}~[cs.LG]
\urldef\tempurl%
\url{https://arxiv.org/abs/2402.12715}
\showURL{%
\tempurl}


\bibitem[Yeh et~al\mbox{.}(2024)]%
        {Yeh:2024:AttentionViz:AGlobalViewofTransformerAttention}
\bibfield{author}{\bibinfo{person}{Catherine Yeh}, \bibinfo{person}{Yida Chen}, \bibinfo{person}{Aoyu Wu}, \bibinfo{person}{Cynthia Chen}, \bibinfo{person}{Fernanda Viégas}, {and} \bibinfo{person}{Martin Wattenberg}.} \bibinfo{year}{2024}\natexlab{}.
\newblock \showarticletitle{AttentionViz: A Global View of Transformer Attention}.
\newblock \bibinfo{journal}{\emph{IEEE Transactions on Visualization and Computer Graphics}} \bibinfo{volume}{30}, \bibinfo{number}{1} (\bibinfo{year}{2024}), \bibinfo{pages}{262--272}.
\newblock
\href{https://doi.org/10.1109/TVCG.2023.3327163}{doi:\nolinkurl{10.1109/TVCG.2023.3327163}}


\bibitem[Yoo and Qi(2021)]%
        {Yoo:2021:TowardsImprovingAdversarialTrainingofNLPModels}
\bibfield{author}{\bibinfo{person}{Jin~Yong Yoo} {and} \bibinfo{person}{Yanjun Qi}.} \bibinfo{year}{2021}\natexlab{}.
\newblock \showarticletitle{Towards Improving Adversarial Training of {NLP} Models}. In \bibinfo{booktitle}{\emph{Findings of the Association for Computational Linguistics: EMNLP 2021}}. \bibinfo{publisher}{Association for Computational Linguistics}, \bibinfo{address}{Punta Cana, Dominican Republic}, \bibinfo{pages}{945--956}.
\newblock
\href{https://doi.org/10.18653/v1/2021.findings-emnlp.81}{doi:\nolinkurl{10.18653/v1/2021.findings-emnlp.81}}


\bibitem[Zhang et~al\mbox{.}(2022)]%
        {Zhang:2022:MachineLearningTestingSurveyLandscapesandHorizons}
\bibfield{author}{\bibinfo{person}{Jie~M. Zhang}, \bibinfo{person}{Mark Harman}, \bibinfo{person}{Lei Ma}, {and} \bibinfo{person}{Yang Liu}.} \bibinfo{year}{2022}\natexlab{}.
\newblock \showarticletitle{Machine Learning Testing: Survey, Landscapes and Horizons}.
\newblock \bibinfo{journal}{\emph{IEEE Transactions on Software Engineering}} \bibinfo{volume}{48}, \bibinfo{number}{1} (\bibinfo{year}{2022}), \bibinfo{pages}{1--36}.
\newblock
\href{https://doi.org/10.1109/TSE.2019.2962027}{doi:\nolinkurl{10.1109/TSE.2019.2962027}}


\bibitem[Zhang et~al\mbox{.}(2020b)]%
        {Zhang:2020:AdversarialAttacksonDeeplearningModelsinNaturalLanguageProcessingASurvey}
\bibfield{author}{\bibinfo{person}{Wei~Emma Zhang}, \bibinfo{person}{Quan~Z. Sheng}, \bibinfo{person}{Ahoud Alhazmi}, {and} \bibinfo{person}{Chenliang Li}.} \bibinfo{year}{2020}\natexlab{b}.
\newblock \showarticletitle{Adversarial Attacks on Deep-learning Models in Natural Language Processing: A Survey}.
\newblock \bibinfo{journal}{\emph{ACM Trans. Intell. Syst. Technol.}} \bibinfo{volume}{11}, \bibinfo{number}{3}, Article \bibinfo{articleno}{24} (\bibinfo{date}{apr} \bibinfo{year}{2020}), \bibinfo{numpages}{41}~pages.
\newblock
\showISSN{2157-6904}
\href{https://doi.org/10.1145/3374217}{doi:\nolinkurl{10.1145/3374217}}


\bibitem[Zhang and Hou(2013)]%
        {Zhang:2013:ExtractingProblematicApiFeaturesFromForumDiscussions}
\bibfield{author}{\bibinfo{person}{Yingying Zhang} {and} \bibinfo{person}{Daqing Hou}.} \bibinfo{year}{2013}\natexlab{}.
\newblock \showarticletitle{Extracting problematic API features from forum discussions}. In \bibinfo{booktitle}{\emph{2013 21st International Conference on Program Comprehension (ICPC)}}. \bibinfo{pages}{142--151}.
\newblock
\href{https://doi.org/10.1109/ICPC.2013.6613842}{doi:\nolinkurl{10.1109/ICPC.2013.6613842}}


\bibitem[Zhang et~al\mbox{.}(2020a)]%
        {Zhang:2020:EffectOfConfidenceAndExplanationOnAccuracyAndTrustCalibrationInAiAssistedDecisionMaking}
\bibfield{author}{\bibinfo{person}{Yunfeng Zhang}, \bibinfo{person}{Q.~Vera Liao}, {and} \bibinfo{person}{Rachel K.~E. Bellamy}.} \bibinfo{year}{2020}\natexlab{a}.
\newblock \showarticletitle{Effect of confidence and explanation on accuracy and trust calibration in AI-assisted decision making}. In \bibinfo{booktitle}{\emph{Fairness, Accountability, and Transparency}} (Barcelona, Spain). \bibinfo{publisher}{Association for Computing Machinery}, \bibinfo{address}{New York, USA}, \bibinfo{pages}{295–305}.
\newblock
\showISBNx{9781450369367}
\href{https://doi.org/10.1145/3351095.3372852}{doi:\nolinkurl{10.1145/3351095.3372852}}


\bibitem[Zhao et~al\mbox{.}(2024)]%
        {Zhao:2024:ExplainabilityforLargeLanguageModelsASurvey}
\bibfield{author}{\bibinfo{person}{Haiyan Zhao}, \bibinfo{person}{Hanjie Chen}, \bibinfo{person}{Fan Yang}, \bibinfo{person}{Ninghao Liu}, \bibinfo{person}{Huiqi Deng}, \bibinfo{person}{Hengyi Cai}, \bibinfo{person}{Shuaiqiang Wang}, \bibinfo{person}{Dawei Yin}, {and} \bibinfo{person}{Mengnan Du}.} \bibinfo{year}{2024}\natexlab{}.
\newblock \showarticletitle{Explainability for Large Language Models: A Survey}.
\newblock \bibinfo{journal}{\emph{ACM Trans. Intell. Syst. Technol.}} \bibinfo{volume}{15}, \bibinfo{number}{2}, Article \bibinfo{articleno}{20} (\bibinfo{date}{feb} \bibinfo{year}{2024}), \bibinfo{numpages}{38}~pages.
\newblock
\showISSN{2157-6904}
\href{https://doi.org/10.1145/3639372}{doi:\nolinkurl{10.1145/3639372}}


\bibitem[Zhou et~al\mbox{.}(2024)]%
        {Zhou:2024:EvaluatingtheValidityofWordLevelAdversarialAttackswithLargeLanguageModels}
\bibfield{author}{\bibinfo{person}{Huichi Zhou}, \bibinfo{person}{Zhaoyang Wang}, \bibinfo{person}{Hongtao Wang}, \bibinfo{person}{Dongping Chen}, \bibinfo{person}{Wenhan Mu}, {and} \bibinfo{person}{Fangyuan Zhang}.} \bibinfo{year}{2024}\natexlab{}.
\newblock \showarticletitle{Evaluating the Validity of Word-level Adversarial Attacks with Large Language Models}. In \bibinfo{booktitle}{\emph{Findings of the Association for Computational Linguistics}}. \bibinfo{address}{Thailand}, \bibinfo{pages}{4902--4922}.
\newblock
\href{https://doi.org/10.18653/v1/2024.findings-acl.292}{doi:\nolinkurl{10.18653/v1/2024.findings-acl.292}}


\bibitem[Zini et~al\mbox{.}(2022)]%
        {ElZini:2022:OntheEvaluationofthePlausibilityandFaithfulnessofSentimentAnalysisExplanations}
\bibfield{author}{\bibinfo{person}{Julia~El Zini}, \bibinfo{person}{Mohamad Mansour}, \bibinfo{person}{Basel Mousi}, {and} \bibinfo{person}{Mariette Awad}.} \bibinfo{year}{2022}\natexlab{}.
\newblock \showarticletitle{On the Evaluation of the Plausibility and Faithfulness of Sentiment Analysis Explanations}. In \bibinfo{booktitle}{\emph{Artificial Intelligence Applications and Innovations}}. \bibinfo{publisher}{Springer International Publishing}, \bibinfo{address}{Cham}, \bibinfo{pages}{338--349}.
\newblock
\showISBNx{978-3-031-08337-2}


\end{thebibliography}










\end{document}